\documentclass[1p]{elsarticle}

\usepackage{lineno,hyperref}
\usepackage{amsmath}
\usepackage{mathtools}
\usepackage{amssymb}
\usepackage{subfig}
\usepackage{soul}

\newcommand{\bl}{\begin{linenomath}}
\newcommand{\el}{\end{linenomath}}
\newcommand{\be}{\begin{equation}}
\newcommand{\ee}{\end{equation}}

\usepackage{xcolor}
\definecolor{backgreen}{rgb}{0.00, 0.169, 0.212}
\definecolor{textgray}{rgb}{0.514, 0.580, 0.589}

\usepackage{lineno,hyperref}
\usepackage{amsmath}

\usepackage{verbatim}
\usepackage{amssymb}

\nolinenumbers

\usepackage{mathtools}

\newcommand{\beq}{\begin{linenomath}\begin{equation}}
\newcommand{\eeq}{\end{equation}\end{linenomath}}
\newcommand{\bseq}{\begin{linenomath}\begin{equation*}}
\newcommand{\eseq}{\end{equation*}\end{linenomath}}
\newcommand{\bln}{\begin{linenomath}}
\newcommand{\eln}{\end{linenomath}}
\newcommand{\bal}{\begin{aligned}}
\newcommand{\eal}{\end{aligned}}
\newcommand{\bgat}{\begin{gathered}}
\newcommand{\egat}{\end{gathered}}

\renewcommand{\bar}{\overline}

\journal{Advances in Water Resources}

\begin{document}


\begin{frontmatter}

\title{On the separate treatment of mixing and spreading by the reactive-particle-tracking algorithm: An example of accurate upscaling of reactive Poiseuille flow\tnoteref{mytitlenote}}
\tnotetext[mytitlenote]{This material is based upon work supported by, or in part by, the US Army Research Office under Contract/Grant number W911NF-18-1-0338. The authors were also supported by the National Science Foundation under awards EAR-1417145, DMS-1211667, DMS-1614586, EAR-1351625, EAR-1417264, EAR-1446236, and CBET-1705770}



\author{David A. Benson\corref{mycorrespondingauthor}}
\address{Hydrologic Science and Engineering, Colorado School of Mines, Golden, CO 80401, USA}

\author{Stephen Pankavich}
\address{Department of Applied Mathematics and Statistics, Colorado School of Mines, Golden, CO, 80401, USA}
\author{Diogo Bolster}
\address{Department of Civil and Environmental Engineering and Earth Sciences, University of Notre Dame, Notre Dame, IN 46556, USA}


\begin{abstract}
The Eulerian advection-dispersion-reaction equation (ADRE) suffers the well-known scale-effect of reduced apparent reaction rates between chemically dissimilar fluids at larger scales (or dimensional averaging).  The dispersion tensor in the ADRE must equally and simultaneously account for both solute mixing and spreading.  Recent reactive-particle-tracking (RPT) algorithms can, by separate mechanisms, simulate 1) smaller-scale mixing by inter-particle mass transfer, and 2) mass spreading by traditional random walks.  To test the supposition that the RPT can accurately track these separate mechanisms, we upscale reactive transport in Hagen-Poiseuille flow between two plates.  The simple upscaled 1-D RPT model with one velocity value, an upscaled Taylor macro-dispersivity, and the local molecular diffusion coefficient matches the results obtained from a detailed 2-D model with fully described velocity and diffusion.  Both models use the same thermodynamic reaction rate, because the rate is not forced to absorb the loss of information upon upscaling.  Analytic and semi-analytic upscaling is also performed using volume averaging and ensemble streamtube techniques.  Volume averaging does not perform as well as the RPT, while ensemble streamtubes (using an effective dispersion coefficient along with macro-dispersion) perform almost exactly the same as RPT.

\end{abstract}


\end{frontmatter}

\section{Introduction}
A recent improvement of the reactive-particle-tracking (RPT) method allows mass transfer between particles and subsequent reactions between any number of chemical constituents on the particles \cite{Benson_arbitrary}.  One of the features of this algorithm is that inter-particle mixing occurs separately from dispersive random walks.  True mixing between dissimilar fluids usually occurs on smaller scales and at slower rates than the dispersive spreading \cite{Danckwerts1953,Nauman1983,Hill1976,Molz_1988,Cirpka2000a,marco_mix_spread,Desimoni2005,donado,Rezaei, Dentz2011,Lehwald2012,tanguy_prl,Tanguy_GRL2014,Ding_WRR}.  {\em Schmidt et al.} \cite{Schmidt_accuracy} suggested that the separate simulation of mixing and spreading by the RPT method could provide a way to accurately upscale reactive solute transport, because the smaller-scale true mixing dictates reaction rates, while the random walks simulate the process of particle separation that accompanies sub-grid (upscaled) velocity perturbations.  Recent work has further extended the particle methods to allow fluid/solid interaction \cite{Schmidt_fluid_solid}.   For additional reasons, such as surface area scaling and solubility saturations near mineral grains, dissolution/precipitation reactions also suffer significant scaling effects of reaction rates (see \cite{White_Peterson_1990,Brantley_book}).

Two of the classic examinations of the disparity between mixing and spreading in moving fluids (and the effect on global reaction rates) were performed by {\em Kapoor et al.} \cite{Kapoor1997,Kapoor1998a}.  These authors chose a simple system because it can be completely defined at the pore scale: laminar miscible displacement of chemically distinct (and reactive) fluids in Hagen-Poiseuille flow, either in a tube or between plates.  In these cases, transport is exactly known, with the well-known parabolic velocity profile between no-slip walls, and random motion solely by molecular diffusion.  The higher velocities at the center of the tube cause overlap of the fluids when projected to 1-D, but mixing only occurs along the warped interface. This system exemplifies the lag of mixing behind spreading in non-uniform velocity fields.

The spreading rate was first derived for Poiseuille flow by Taylor \cite{Taylor1953}, who showed that the 2-D transport of nonreactive tracer in a tube could be upscaled (averaged) to 1-D.  Given enough time to sample the entire velocity variability by local diffusion, transport can be effectively described by a one-dimensional advection-dispersion equation with constant velocity and an enhanced macro-dispersion coefficient reflecting subscale advection-induced spreading.  The asymptotic ($t\rightarrow \infty$) upscaled longitudinal hydrodynamic macro-dispersion coefficient $D_{mac}$ may be orders-of-magnitude larger than the local-scale molecular diffusion coefficient $D_{mol}$, and its functional form depends on the shear velocity distribution and molecular diffusion coefficient \cite[e.g.,][]{Bolster2010}. 

This has inspired two tacks for upscaled reactive transport.  The first tack has derived two dispersion coefficients, one for the effective mixing $D_{\mathit{eff}}$ and another corresponding to Taylor's $D_{mac}$ that describes macro-dispersive spreading.  In this approach, mixing is assumed to be the dominant mechanism dictating reaction rates (i.e., reactions are nearly instantaneous) so that the statistics of mixing via destruction of concentration gradients give an effective, smaller, dispersion coefficient \cite{Dentz2000,Cirpka2000a,Cirpka2000b,marco_mix_spread}.  This smaller $D_{\mathit{eff}}$ is designed to slow down reactions, but not place solutes in the correct locations, so \cite{Cirpka2000b} suggests a streamtube approach in which the spreading of the centers of mass for the streamtubes is given by $D_{mac}$, while mixing within a streamtube is given by the smaller $D_{\mathit{eff}}$. The second tack seeks to adjust the reaction rate itself by recognizing that the reactant segregation that results from upscaling should modify the effective reaction rate.  This approach has been used on simpler (diffusion-only) problems that allow direct calculation of segregation---as measured by reactant concentration covariance evolution equations \cite{Bolster2012AWR,Tartakovsky2012,Paster_JCP,Schmidt2017}.   However, in heterogeneous velocity distributions, these equations have yet to be analytically solved, and only simpler expressions based on very fast or very slow reactions end-members have been developed \cite{Porta2012}.


These upscaling approaches point out the fundamental problem associated with an Eulerian advection, dispersion, and reaction equation (ADRE).  The dispersion coefficient valid for conservative transport at some scale will over-predict fluid mixing and reaction {at the same scale, but under-predict spreading at some larger scale} (or volume averaged to fewer spatial dimensions).  We seek to correct this problem with a Lagrangian framework.  For $M$ species undergoing Fickian dispersion in incompressible flow, the coupled ADREs are
\bl
\be \label{eq:ADRE}
\frac {\partial c_i} {\partial t}=- {\bf{v}}\cdot \nabla c_i +\nabla \cdot \left ({\bf D} \nabla c_i \right) +R(c_A,c_B,...,c_M, k_{1}, ..., k_{N});\quad\quad i=A,B,...,M
\ee
\el
\noindent where $c_i$ is the concentration of each of the species labeled $i=A,B,..., M$, ${\bf v}$ is a velocity vector, ${\bf D}$ is a dispersion tensor, and $R()$ is a reaction function among the $M$ constituents with $N$ reaction channels.  The ADRE assumes that ${\bf D}$ describes mixing and spreading in exactly the same way.  For continuously varying ${\bf v}$, this is only true at the molecular scale.  In practice, however, all variables and parameters in \eqref{eq:ADRE} have some finite support scale, and the discrepancy between mixing and spreading grows with support scale \cite{Dentz2011,tanguy_prl,Tanguy_GRL2014}.  Indeed, if $\bf{v}$ is given by Darcy's Law and a hydraulic conductivity parameter, then this type of upscaling has already occurred.  {As briefly reviewed above,} this discrepancy may be accounted for by adjusting the only remaining equation parameters that are held in the reaction term $R()$, or by solving the equation separately with larger and smaller $\bf{D}$ to figure out mixing versus proper positions of reactants.  If the perturbations of $c_i, {\bf v}$, and ${\bf D}$ are known, as well as their auto- and cross-correlations in time and space, then the adjusted $R()$ can be approximated with closure assumptions \cite{Dentz2011,Tartakovsky2012,Porta2012,Porta2013} that may not be particularly accurate {for some values of coefficients. We include a brief comparison of the two most notable analytic upscaling approximations to our numerical method in this paper.}

On the other hand, the micro-scale physics of particle motion and interaction may already carry all information neglected by the analytic upscaling.  Here we show that, for the simplest case, the RPT method does indeed automatically track the necessary small-scale information and performs a natural upscaling.


\section{Hagen-Poiseuille Flow}\label{sec_HP}

\begin{figure}[t]
 \centering
 \includegraphics[width=1.1\textwidth]{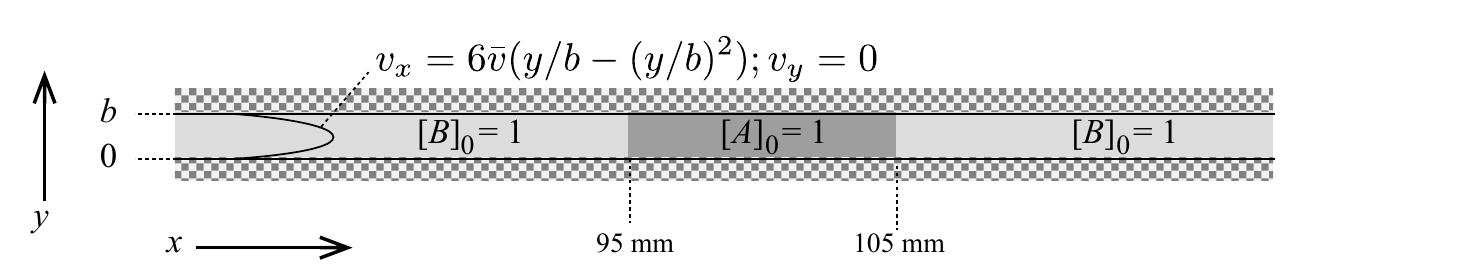}
 \caption{Schematic of physical setup and initial condition.  Not to scale.} 
 \label{fig:schematic}
 \end{figure}

We simulate an identical problem of flow, transport, and kinetic bimolecular reaction $A+B\rightarrow C$ between two parallel plates as did {\em Kapoor et al.} \cite{Kapoor1997} (Figure \ref{fig:schematic}). The concentration units are arbitrary, but we will use moles/L (molar).  {The local thermodynamic reaction rate is given by the law of mass action $R=-kc_Ac_B$ for constituents $A$ and $B$, where $k$ [molar$^{-1}$ T$^{-1}$] is a rate coefficient. Without loss of generality we assume unit activity coefficients.  This type of reaction has been studied experimentally and theoretically because of its simplicity and dependence on local mixing (e.g., \cite{Raje,gramling,Edery2009, Edery2010, sanchez-vila, Zhang_PRE_react,dong_awr}).}  The plates are separated by aperture $b=1$ mm, and the molecular diffusion coefficient is $10^{-3}$ mm$^2$/s.  An initial slug of reactant $A$ is placed across the entire aperture from  $x=95$ to $x=105$ mm (zero elsewhere), while reactant $B$ is placed (with unlimited extent) only on either side of the slug.  The velocity field follows $ {\bf{v}} =[ v_x(y), 0 ]$, where $v_x(y) = 6\overline{v} \left ( \frac{y}{b} - \left ( \frac{y}{b} \right)^2 \right )$.  {The mean velocity is specified as $\bar v = 1.0132$ mm/s, giving a characteristic advection time of  $t_a=b/\bar v =0.99$ s.  {\em Kapoor et al.} define an effective diffusion time $t_D=(b/\pi)^2/D_{mol}$ different from more recent definitions of $t_D=b^2/D_{mol}$ (e.g., \cite{Porta2012}), so their Peclet and Damkohler numbers (100 and 10, respectively) are ``off'' by a factor of $\pi^2\approx 10$.  Our Peclet number is defined by $t_D/t_a = \bar v b / D_{mol} = 1013$.  {\em Kapoor et al.} chose a reaction rate coefficient $k=0.0987$ (Mol s)$^{-1}$ to yield a Damkohler number of $Da =  kc_A(t=0) b^2/D_{mol} = 98.7$, where the initial nonzero reactant concentrations are $c_A=c_B=1$ molar. }

\begin{figure}[h!]
 \centering
 \includegraphics[width=0.99\textwidth]{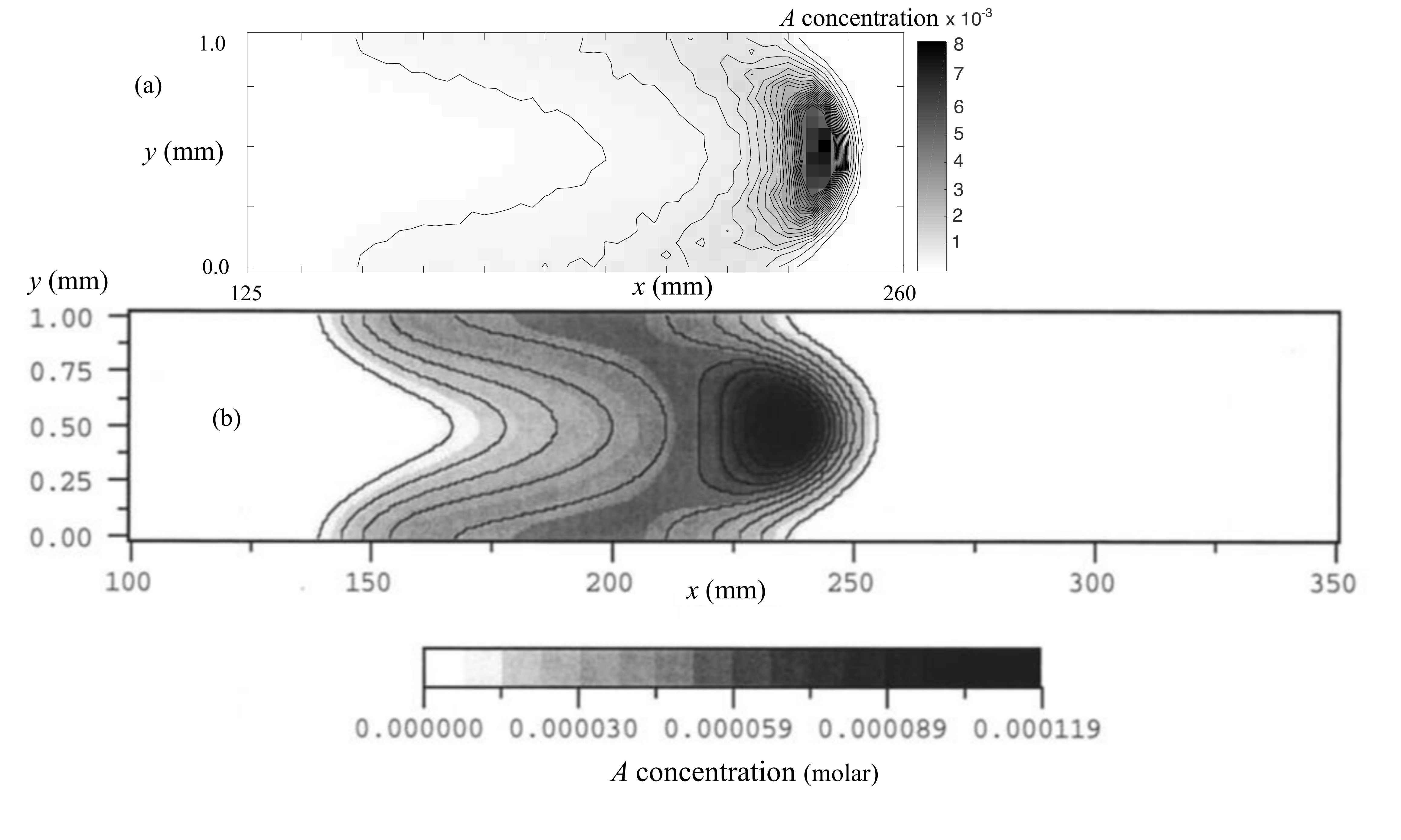}
 \caption{2-D maps of reactant $A$ concentration after $t_D=0.1$ ($t=100$ s): a) Using RPT method \cite{Bolster_mass}, and b) using finite-differences (reproduced from \cite{Kapoor1997}).  Initial slug of $A$ at unit concentration placed at $95\le x\le 105$ mm. Contour interval in (a) is $2.5 \times 10^{-4}$ molar.  Note the sharper gradients and much higher concentrations (maximum approximately 70 times) maintained by RPT method.} 
 \label{fig:A_map}
 \end{figure}

First, we construct a 2-D simulation of the system using the particle-number-preserving method \cite{Bolster_mass}. {This is an extension of the original RPT algorithm that killed entire particles on reaction (see \cite{Benson_react}).  The newer algorithm makes each particle carry only one species, and the mass and concentration on that particle is continuously adjusted to account for reactions.}  We chose this method because the reactions are independent of the particle random walks, and we need to accurately track the variance of particle positions to validate upscaling to 1-D.  Increasing particle numbers until convergence showed that 20,000 $A$ particles and 240,000 total $B$ particles in two 60 mm zones on either side of the initial $A$ slug were sufficient.  A gray-scale plot of the binned concentrations of $A$ remaining at $t_D=0.1$ ($t =  100$ s) shows the segregation of reactants that results from the parabolic velocity profile (Figure \ref{fig:A_map}a).  {\em Kapoor et al.} used centered finite-differences in their solution, with a maximum grid Peclet number of $v_{max}\Delta x/D_{mol} = 37.5$, so that their solution was vastly artificially over-mixed (and over-reacted).  Their peak concentrations of remaining reactant $A$ are about 70 times less than ours, but the general shapes agree quite well. If one wished to use a {first-order accurate (similar to {\em Kapoor et al.} \cite{Kapoor1997}) } finite-difference method with a grid Peclet number on the order of unity, then $\Delta x \approx 1.7 \times 10^{-4}$. Using 100 nodes in the $y$-direction, the 200 mm $\times$ 1 mm domain would require on the order of 120 million nodes.  {A more accurate advection scheme would require fewer nodes but would still be computationally demanding so we use the particle method for the benchmark 2-D simulations (see \cite{Benson_AWR_2016} for a comparison of methods)}.

\begin{figure}[t]
 \centering
 \includegraphics[width=0.9\textwidth]{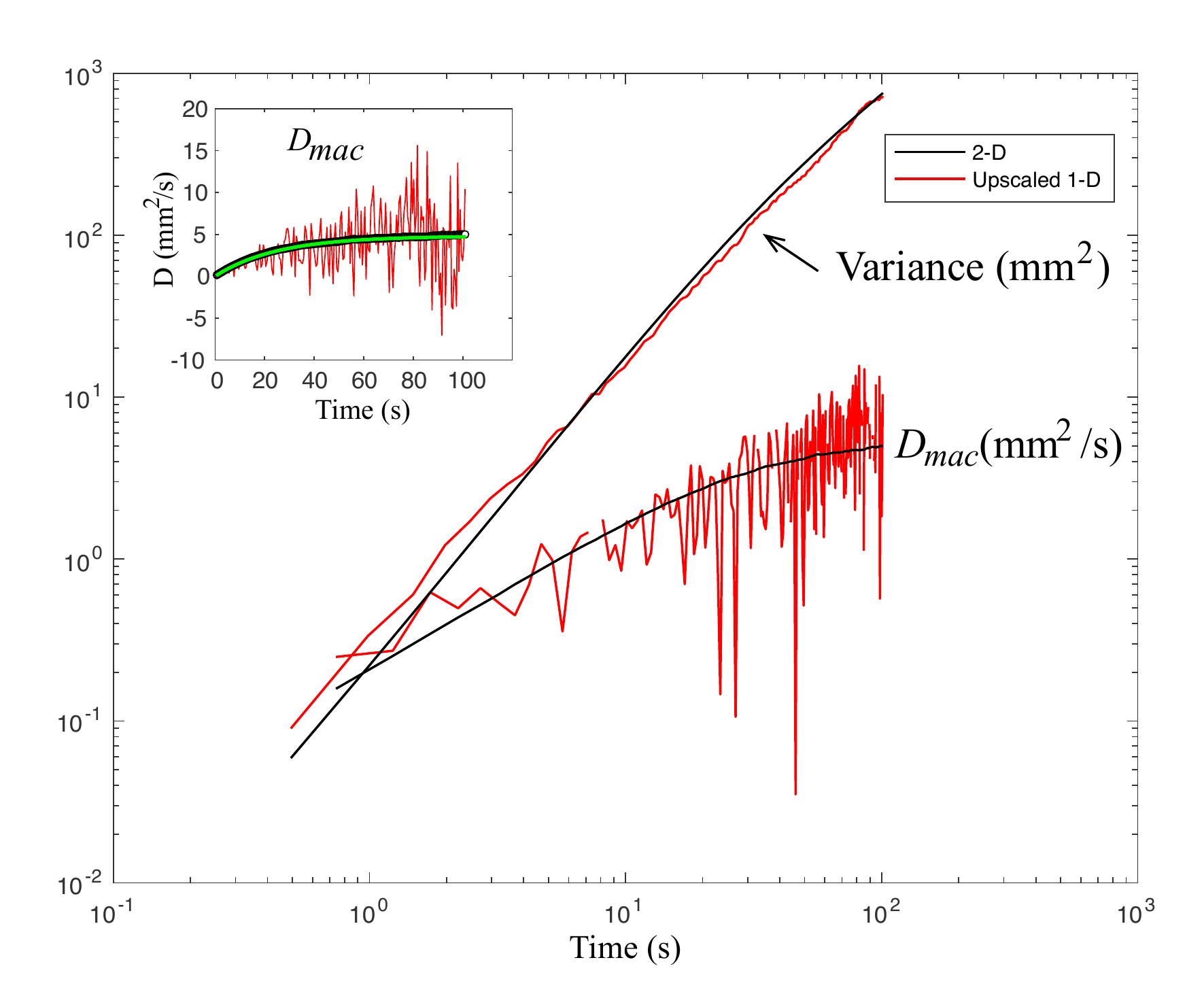}
 \caption{Log-log plot of $A$ particle variance and $D_{mac}$ estimation from both 2-D and upscaled 1-D models.  The exponential model (green curve, inset) is shown against 2-D model (red) and 1-D posterior model results using the exponential.  The 1-D model variance and posterior $D$ show noise due to the smaller number of $A$ particles (447).} 
 \label{fig:var}
 \end{figure}

The RPT algorithm of \cite{Bolster_mass} was used because individual particles are composed solely of $A$, $B$, or $C$.  This allows us to track all $A$ particles in the initial slug to verify the analytically upscaled value of $D_{mac}(t)$ (derived in the Appendix). The centered second moment (i.e., the ``plume variance") grows as expected (Figure \ref{fig:var}): quadratically near $t=0$ because of ballistic particle motion according to the velocity profile, transitioning to linear growth according to Fick's law. We took first differences in discretized time to calculate $D_{mac}(t) \approx \Delta \text{VAR}(X_A)/(2\Delta t)$.  We fit $D_{mac}(t)=4.88\frac{mm^2}{s}[1-\exp(-t/25s)]$ for the upscaled (1-D) model from the 2-D data. Note that the analytic $D_{mac}(t)$ is almost exactly equal to our measured $D_{mac} (t)$ (see Appendix).

The locally well-mixed 2-D model used a particle density of 2,000/mm$^2$, or a 2-D volume per particle of $5\times10^{-4}$ mm$^2$. An equivalent volume, or average spacing, in 1-D is defined by $\Delta s = (5\times 10^{-4})^{1/2}=0.0224$ mm.  Therefore, for the 10 mm initial condition for reactant $A$, an equivalently well-mixed 1-D model would use $N=10$ mm/0.0224 mm = 447. The total particle number including the surrounding initial $B$ reactant is 5811.   Upon upscaling, however, the concentrations are not locally well-mixed.  The projection to 1-D places disparate concentrations in the same $x$-location.  The particles may represent these different concentrations, and do so by representing some support volume.  This volume has been shown \cite{Paster_JCP,Schmidt2017} to represent the concentration auto-correlation (and as time grows, the cross-correlation) distance.  {Previous studies (e.g., \cite{Kapoor1997}) have shown that, in Poiseuille flow, the reactants segregate in regions that occupy about one-half the width of the aperture.  This is verified in Figure \ref{fig:A_map}a.  
Therefore, each particle occupies a 1-D volume upon upscaling of $\approx0.5$ mm.  This gives an initial particle number of 20 for the $A$ initial condition, or 260 total particles. {This visual estimate is formally shown to be representative in the Appendix}. These particle numbers (5811 and 260, for well-mixed and segregated)} were placed in a 1-D model using a newer particle-tracking algorithm \cite{Benson_arbitrary,Schmidt_accuracy}.  For the lower number, we also used an ensemble of 100 simulations due to noisy runs. 

This newer algorithm \cite{Benson_arbitrary,Schmidt_accuracy} transfers masses or moles of any and all species ({denoted by a superscipt, i.e.,} species $A$ as $m^A$) during a time-step of duration $\Delta t$ between all particle pairs $i$ and $j$ {(denoted by subscripts)} according to 
\bl
\be \label{eq:transfer}
m_j^A(t+\Delta t) = m_j^A(t) + \sum_{i} \frac{1}{2}(m_i^A(t)-m_j^A(t))P_{i,j},
\ee
\el
using each particle pair's collocation probability 
\bl
\be \label{eq:v(s)}
P_{i,j}= (\Delta s / \sqrt{8\pi \eta D_{mol} \Delta t}) \exp(-r^2/(8\eta D_{mol}\Delta t)),
\ee
\el
where $\Delta s$ is the particle support volume, $r$ is the distance between the $i$ and $j$ particles, {and $0<\eta<1$ is the fraction of the isotropic diffusion simulated by interparticle mass transfer.  Because the mass transfer process is diffusive \cite{Schmidt_accuracy}, any ``leftover'' diffusion is added to macro-dispersion, i.e., by $(1-\eta)D_{mol}+D_{mac}$, which is simulated by random walks \cite{Benson_arbitrary}.  We chose $\eta=1/2$, although it makes no observable difference as long as the value is not extremely close to 0 or 1. }   {Note that $\Delta s$ is essentially calculated automatically as for any $j$, the probabilities must satisfy $\sum_i P_{i,j} = 1$; therefore, a matrix of probabilities is adjusted to have a column sum of unity (see \cite{Schmidt_accuracy}).}  Also, the masses in the sum can be from the beginning or end of a time-step, or updated sequentially \cite{Schmidt_accuracy}.  We chose the sequential updating here for stability. After mass transfer of $A$ and $B$ among all particles, the bimolecular reaction proceeds according to a first-order implementation of the law of mass action \cite{Benson_arbitrary}, namely 
\bl
\be \label{eq:rxn}
\Delta m_j^A=\Delta m_j^B=-\Delta m_j^C=-k\Delta t \Delta s (m_j^A/\Delta s)(m_j^B/\Delta s).
\ee
\el

After mass transfer and reaction, the particles experience advection and any additional local and macro-dispersion by well-known random walk methods \cite{labolle,Salamon_2006}. {Specifically, in the 2-D model, particles are advected by $v_x(y)\Delta t$ and isotropically diffused by $(1-\eta)D_{mol}$, while the 1-D models advect by $\bar{v}\Delta t$ and disperse using Taylor's $(1-\eta)D_{mol}+D_{mac}(t)$ unless otherwise specified for hypothesis testing.} 

\section{Results and Discussion}
\begin{figure}
 \centering
 \includegraphics[width=0.85\textwidth]{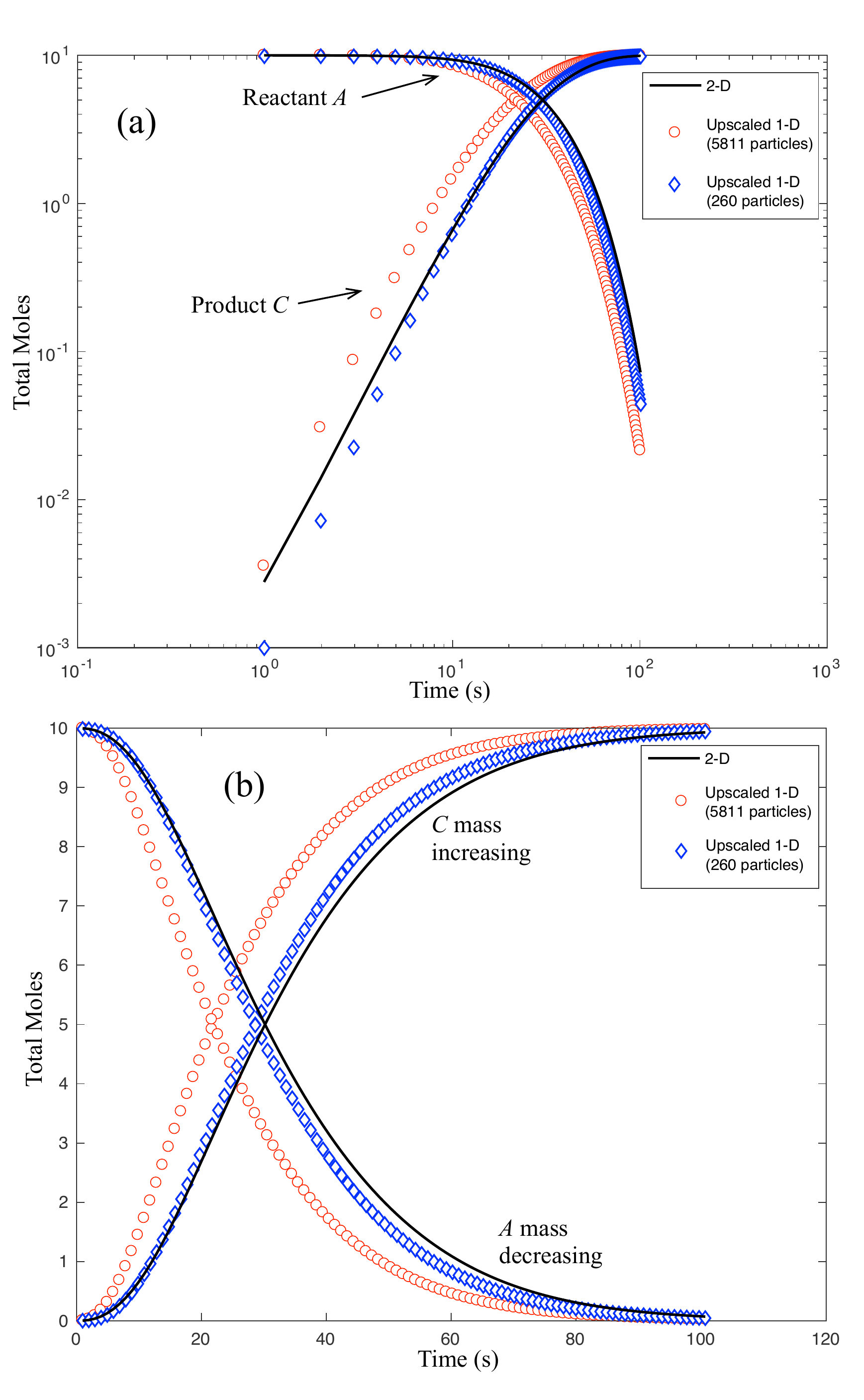}
 \caption{Log-log (a) and linear (b) plots of total domain moles of reactant $A$ and product $C$ in both the 2-D full velocity model (red curves) and 1-D upscaled (single velocity) model (blue circles).  Upscaled 1-D model uses local $D_{mol}=10^{-3}$ mm$^2$/s for inter-particle mixing and Taylor hydrodynamic dispersion $D_{mac}=D_{\mathit{eff}}(t)=4.8(1-\exp(-t/25s))$ mm$^2$/s for random-walk particle spreading.} 
 \label{fig:A_C_mass}
 \end{figure}

The rates of $C$ production and late-time $A$ decline agree quite well in both log-log and linear coordinates, when the lower number of initial $A$ particles ($N=260$) is used to represent concentration fluctuation distances on the order of one-half pore width (Figure \ref{fig:A_C_mass}).  We expect that an even better fit could be achieved by adjusting the particle number, but we have not changed our original visual estimate of particle support volume equal to one-half pore width (see also the appendix).  This level of agreement was not expected due to the loss of detailed velocity information upon upscaling.  
By virtue of a constant velocity and Gaussian random walks, the 1-D model has a Gaussian shape at any time.  The fact that the shape of the ``plume'' appears to be of secondary importance would indicate that the magnitude of concentration fluctuations, along with the rate at which these fluctuations mix at the local scale, is a primary driver of reaction rate (see also \cite{Tartakovsky_Hagan}).  

\begin{figure}[h!]
 \centering
 \includegraphics[width=0.9\textwidth]{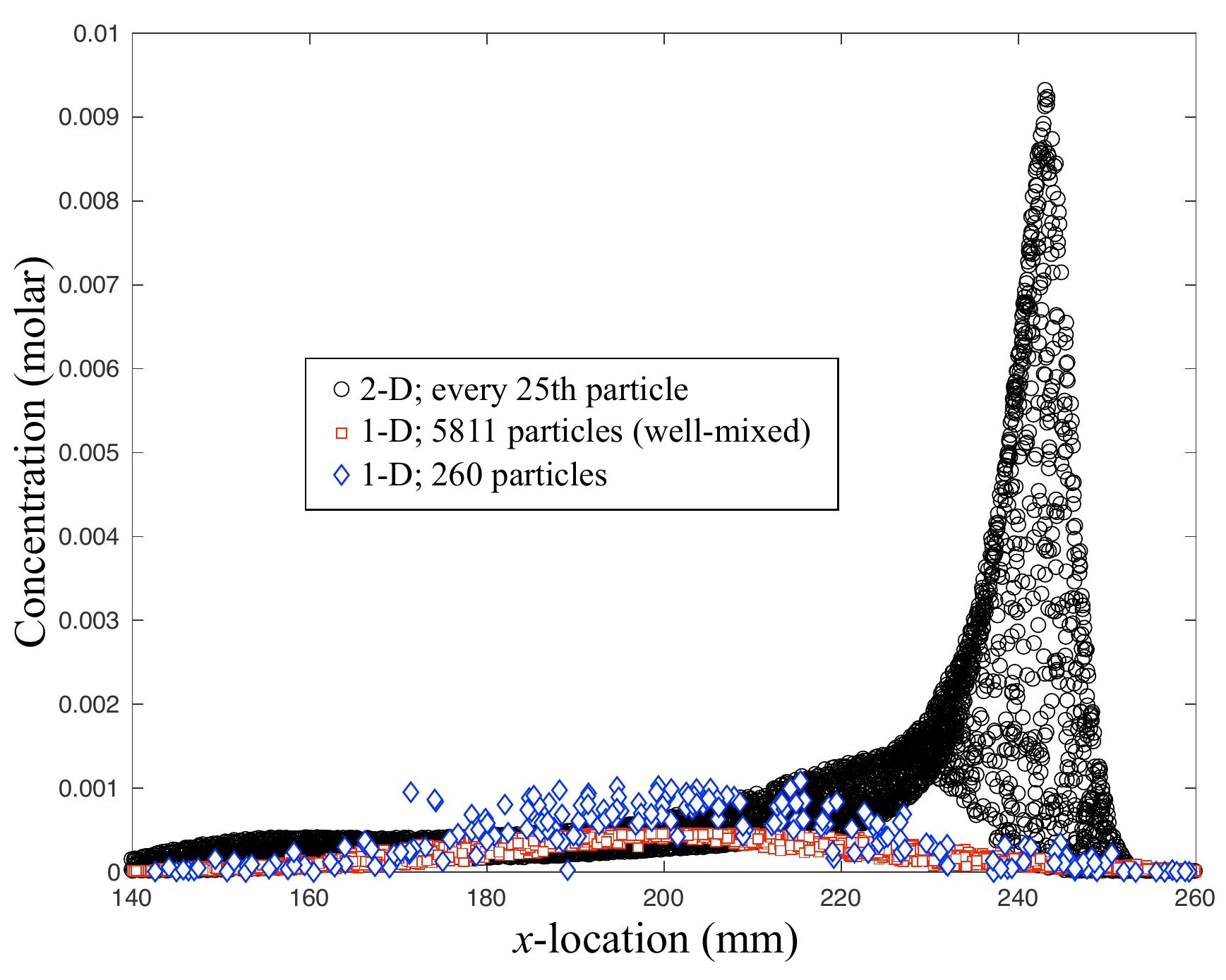}
 \caption{{Particle concentrations at a time of $t_D=b^2/D_{mol}=0.1$ (= 100 s) for single realizations of 1) a 260,000-particle 2-D model (black circles); 2) a 5,811-particle upscaled 1-D model (red squares); and, 3) a 260-particle upscaled 1-D model (blue diamonds). }} 
 \label{fig:mass_v_x}
 \end{figure}

We can inspect the degree to which different concentrations coexist in close proximity by examining a plot of each particle's concentration of $A$ versus position along the $x$-axis in both the 2-D and 1-D RPT models (Figure \ref{fig:mass_v_x}).  For clarity we plot every 25$^{th}$ particle from the 260,000 initial $A$ particle model in 2-D, { along with single realizations from the 5811 particle and 260 particle models in 1-D, all }at a time of 100 s (=$t_D/10$).  Unlike an Eulerian model, the 1-D upscaled models have particles at, or near, the same position with very different concentrations.   This happens because the properly upscaled dispersive random walks take particles with different masses and move them relatively large distances, using $D_{mac}(t)$, into areas of very different {concentrations}.  The local transfer of mass between particles takes place more slowly than this (according to $D_{mol}$), so there is not enough time to equilibrate with surroundings before a new excursion.  {This is especially true in the 260-particle model, and shows that a} key to this upscaling is a correct calculation of the large-scale excursion lengths versus the local-scale mixing rate.  Also evident in Figure \ref{fig:mass_v_x}  is that the 1-D model has lost information about the particular velocity distribution and performs Gaussian random walks.

\begin{figure}
\vspace{-0.5in}
 \centering
 \includegraphics[width=0.85\textwidth]{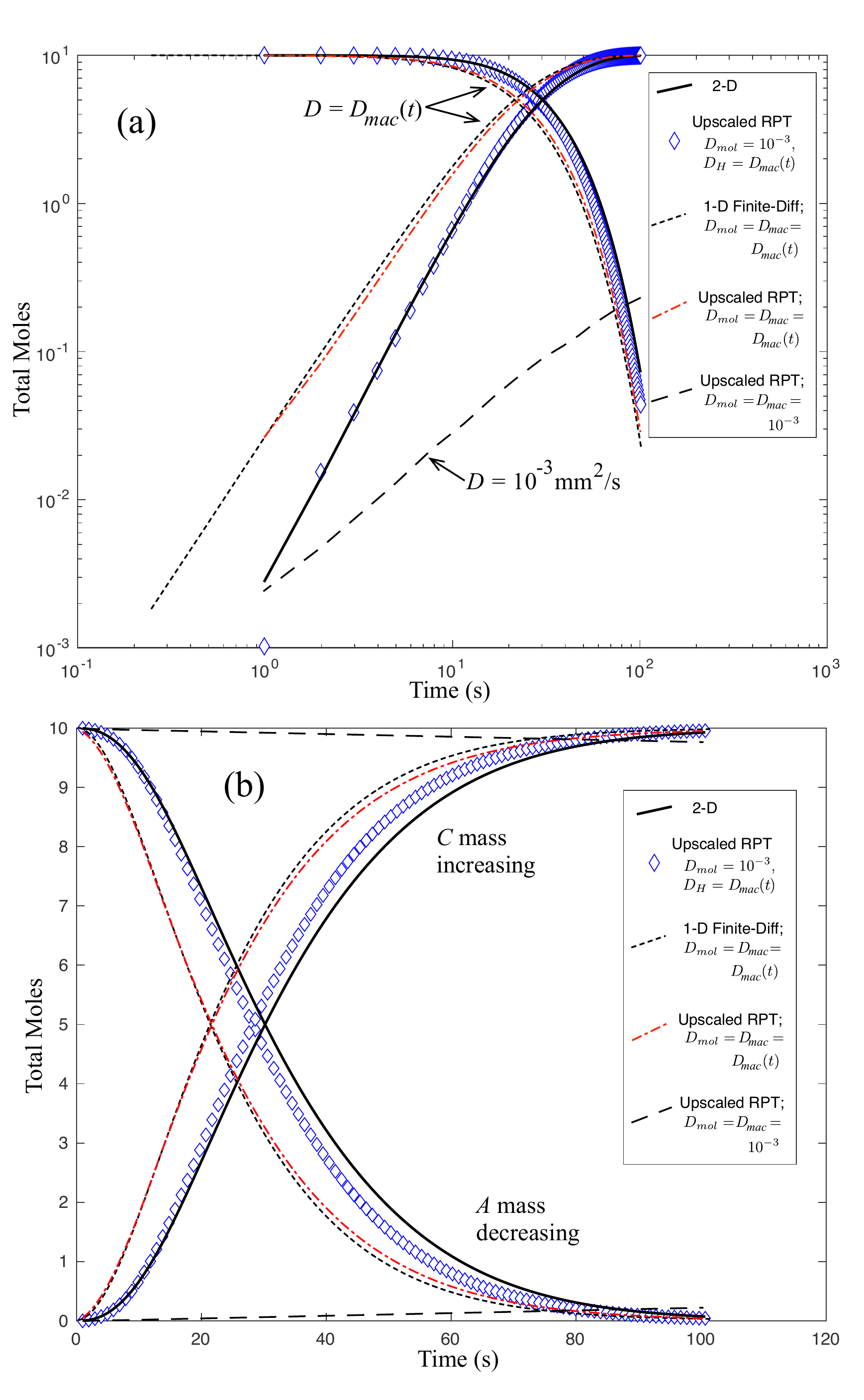}
 \caption{Log-log (a) and linear (b) plots of total domain moles of reactant $A$ (decreasing) and product $C$ (increasing) in both the 2-D full velocity model (black solid curves) and the 1-D upscaled (single velocity) models.  Blue diamond symbols are the previous model using local $D_{mol}=10^{-3}$ mm$^2$/s for inter-particle mixing and Taylor hydrodynamic dispersion $D_{mac}=D_{\mathit{eff}}(t)=4.88(1-\exp(-t/25s))$ mm$^2$/s for random-walk particle spreading. Black dashed (finite-difference model) and red dot-dash (RPT model) curves use $D_{\mathit{eff}}(t)$ for both mixing and spreading, black wide dashed curves use $D_{mol}$ for both in RPT model.} 
 \label{fig:diff_D}
 \end{figure}

\begin{figure}[h!]
\vspace{-0.5in}
 \centering
 \includegraphics[width=0.85\textwidth]{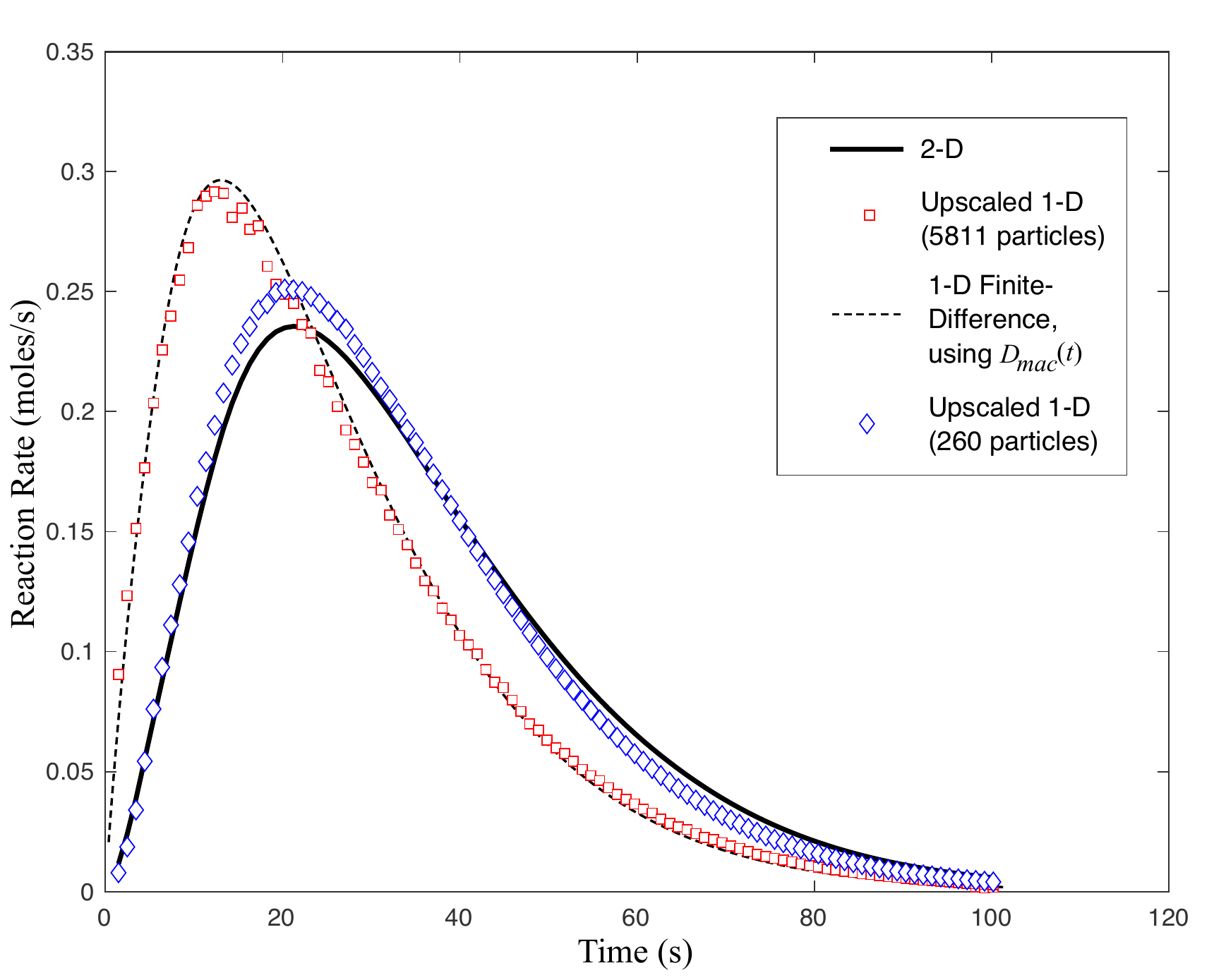}
 \caption{Total reaction rates in full 2-D simulation (solid black curve) and upscaled 1-D models.  The dashed black line is a finite-difference model with $D=D_{\mathit{eff}}(t)$, the red squares represent an RPT model with 447 initial $A$ (5811 total) particles, and the blue diamonds use 20 initial $A$ (260 total) particles.  } 
 \label{fig:reaction_rates}
 \end{figure}

A more important point is the difference between the {Lagrangian} model with mixing given by $D=10^{-3}$ mm$^2$/s and spreading by $D_{mac}(t)$, and {an Eulerian model} that uses the same value (either $D_{mol}$ or $D_{mac}(t)$) for both mixing and spreading. We ran the 1-D models using these potential end-member choices for mixing and dispersion, and found that those upscaled models under-predict and over-predict reaction magnitudes significantly (Figure \ref{fig:diff_D}).  This is further emphasized by a plot of the global reaction rates (Fig. \ref{fig:reaction_rates}).  To check the accuracy of these particle models, we also coded a 1-D Eulerian model using upwind finite-differences (FD) and $D_{mac}(t)$.  With a constant velocity and Courant number of unity, the FD model does not suffer numerical dispersion, and it verifies the RPT model using upscaled {single valued} $D=D_{mac}(t)$.  This model using $D_{mac}(t)$ for both mixing and spreading is a better model than using $D_{mol}$ for both, which indicates that it is more important for the reactants to be placed in the proper {\em positions} before mixing begins, even if the local mixing is overdone.  That is why, for $t\rightarrow 0$, the reactant initial condition specifies the correct positions and using $D_{mol}$ is the better model.  After a time of approximately $t_D/10$, the solute, if placed in the proper positions, becomes better-mixed and using $D_{mac}$ is more appropriate (see \cite{Cirpka2000b,marco_mix_spread,Porta2012}), if one is forced to use a {deterministic} Eulerian model.   

\subsection{Analytic/Semi-analytic Upscaling}

As discussed in the Introduction, two forms of upscaling have been derived and/or suggested for the mixing versus spreading problem.  One extends the approach of {\em Porta et al.} \cite{Porta2012}, who volume-average the reactive transport equations.  Those authors solve for the asymptotic ($t \rightarrow \infty$) coefficients in upscaled equations that keep first-order terms only.  We extend their approach for coefficients that are functions of time (Appendix).  The result is two coupled transport equations, one for a reactive species and another for the conservative species $A-B$.

\begin{figure}
 \centering
 \includegraphics[width=0.85\textwidth]{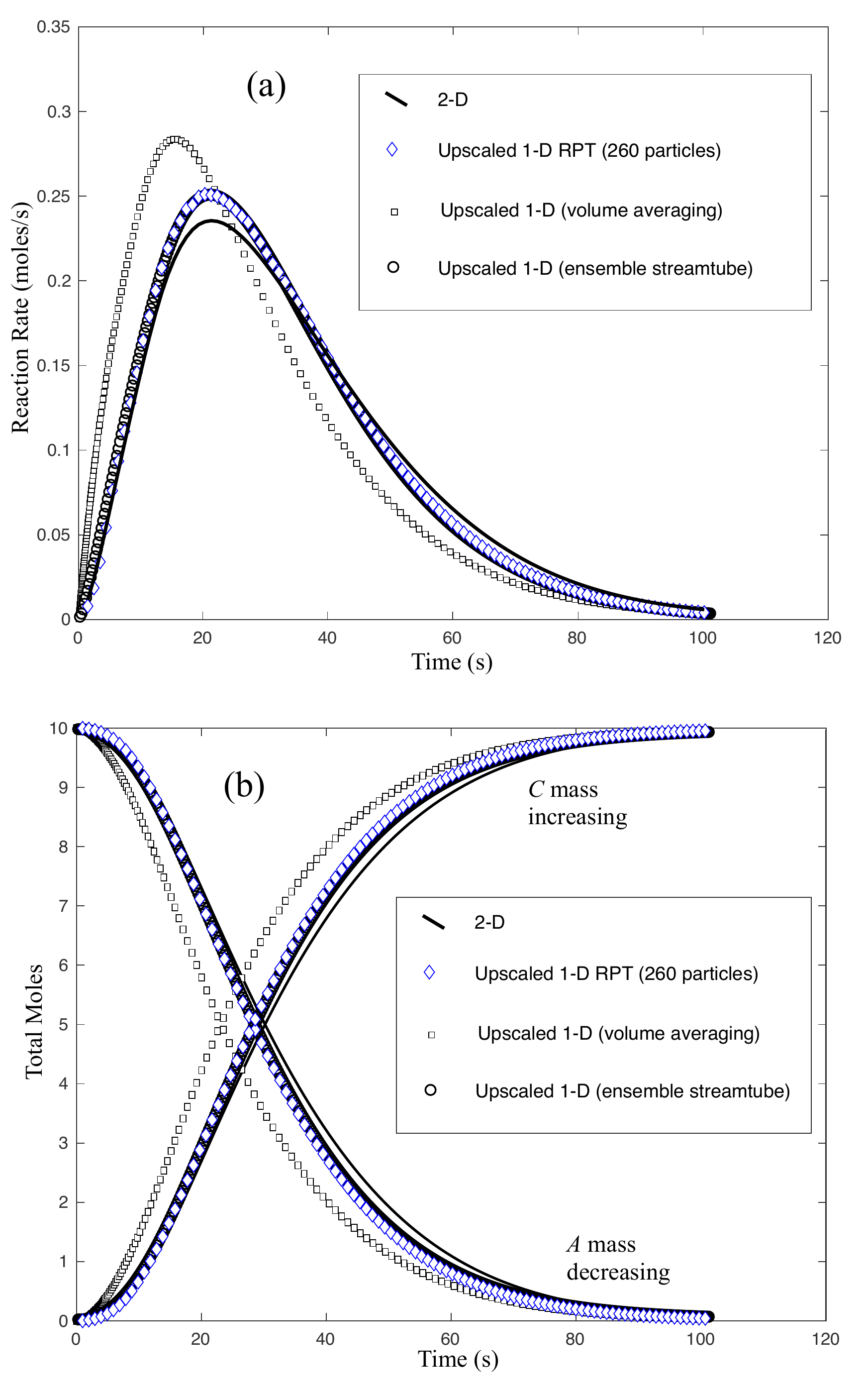}
 \caption{{Results from the simulated upscaled 1-D equations: (a) Total reaction rates and (b) Total simulated domain moles of reactant $A$ and product $C$.  Black squares are volume-averaged equations \eqref{eq:Porta_start} through \eqref{eq:Porta_end}.  Black circles are stream-tube model using $D_{\mathit{eff}}(t)$ in \eqref{eq:ADRE}.  Previously plotted results for the full 2-D and 260 initial $A$-particle 1-D models are included for reference.  The streamtube model results are almost completely obscured by the 260 particle model results.}} 
 \label{fig:Porta}
 \end{figure}

A Galilean-invariant (i.e., $\overline{v}=0$) set of these equations was solved on a finite-difference grid using centered differences for first derivatives, a classical 3-point stencil for second derivatives, and operator-splitting for reaction, so that the solution is $\mathcal{O}(\Delta x^2,\Delta t)$.  Discretization was decreased until convergence was observed with a final $\Delta x =0.1$ mm, $\Delta t =0.001$ s.  Plots of the global reaction rate, and the masses of $A$ and $C$, over time (Fig. \ref{fig:Porta}) show that the perturbation upscaling overpredicts the mixing and reaction rates, although the inclusion of an adjusted reaction rate in \eqref{eq:upscaledB} improves the results relative to a simple 1-D finite difference solution of \eqref{eq:ADRE} using upscaled $D_{mac}(t)$ (compare Fig. \ref{fig:Porta} to Figs. \ref{fig:diff_D} and \ref{fig:reaction_rates}).  The full perturbation-upscaled solution shown in Fig. \ref{fig:Porta}  still overpredicts reaction rates relative to the RPT solution either because of the neglect of higher-order moments or other terms to achieve closure.  Specifically, the regime in question is defined by a moderately fast reaction rate ($Da=10$ as defined by {\em Kapoor et al.} or $Da\approx 100$ by {\em Porta's} definition), which is assumed to be infinite in order for \eqref{eq:B'} to reduce to \eqref{eq:fast}.

\begin{figure}
 \centering
 \includegraphics[width=1.0\textwidth]{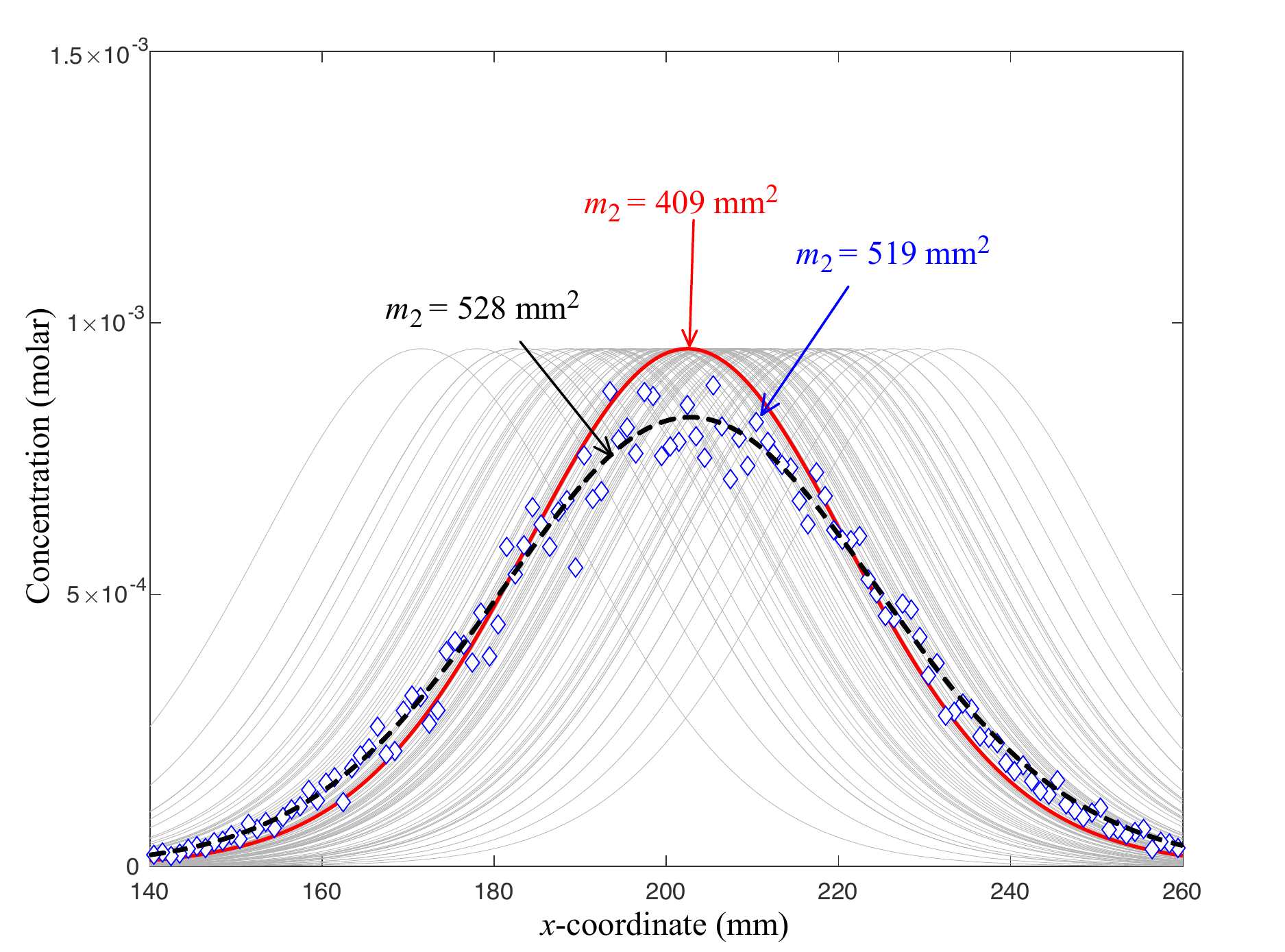}
 \caption{{Concentrations of reactant $A$ versus $x$-coordinate at time $t_D/10 \approx 100$ s in upscaled models: Solid red curve is 1-D finite-difference solutions of \eqref{eq:ADRE} for $i=A,B,C$ using ${\bf D} = D_{\mathit{eff}}(t)$.  Light grey curves are 100 random placements of the red curve with spatial variance $2t\int_0^t [D_{mac}(\tau)-D_{\mathit{eff}}(\tau)]d\tau$.  The black dashed line is the superposition (ensemble average) of 1000 randomly placed grey curves.  Blue diamond symbols are 100-realization ensemble average of the 260-particle RPT model.  Calculated spatial centered second moments of model results (denoted $m_2$) show the close correspondence of the ensemble streamtube and ensemble particle results.}}
 \label{fig:Cirpka}
 \end{figure}

{A second method (that we call here ``ensemble streamtube'') derives two upscaled dispersion coefficients for a conservative tracer: one for effective mixing within a streamtube ($D_{\mathit{eff}}$) and another for the spreading that includes the separation of the centers of mass from different velocities among the streamtubes ($D_{mac}$) \cite{Dentz2000,Cirpka2000a,Cirpka2000b,marco_mix_spread}.  The streamtubes allow mass transfer between each other; this process promotes mixing.  A practical application of this method requires solving the reactive system of transport equations \eqref{eq:ADRE} using $D_{\mathit{eff}}$ and mapping those solutions to the positions of the streamtubes.  At time $t$, the centers of mass have advected to a mean position of $\bar{v}t$ and accumulated an extra (spreading) dispersion coefficient $D_{\Delta}(t)= \int_0^t [D_{mac}(\tau)-D_{\mathit{eff}}(\tau)] d\tau $.  Calculation of $D_{mac}(t)$ is already provided in the Appendix, and using the methodology in \cite{marco_mix_spread} we further calculate $(D_{mac}-D_{\mathit{eff}})$ as a function of time.  The 1-D solution of the reactive system using $D_{\mathit{eff}}(t)$ has an excellent representation of the reaction rate and evolution of reactant and product moles (Fig. \ref{fig:Porta}). A superposition of these 1-D solutions with random mean positions given by $\bar v t + \sqrt{2tD_{\Delta}(t)} \mathcal{Z}$, where $\mathcal{Z}$ is a standard Normal yields plots very close to the ensemble RPT model using 260 particles (Fig. \ref{fig:Cirpka}).  At the time shown in Figure \ref{fig:Cirpka} ($t=100$ s $=t_D/10$), the total moles of $A$ remaining in the ensemble streamtube and ensemble RPT models are 0.0466 and 0.0462, while the centered second moments are 528 and 519 mm$^2$, respectively.  We speculate that, for this moderately fast reaction ($Da=100$), the particle method is a stochastic implementation of the ensemble streamtube method: each particle moves by mean advection and macrodispersion, but transfers mass according to 1) the local diffusive Green's function \cite{Schmidt_accuracy}, and 2) the covariance of concentrations given by the particle support volume. An open question is the regions of the $Da, Pe$ and chemical sequestration parameter space under which the correspondence holds. }
 
\section{Conclusions}
In this technical note we show that the RPT method can accurately simulate dimensionally upscaled transport and reaction for pre-asymptotic times 
in Poiseuille flow.  We used the time-dependent, upscaled effective Taylor macro-dispersion coefficient $D_{mac}$ for random walks and the isotropic molecular diffusion for locally diffusive mass transfer between particles.  An accurate model using finite differences would require full specification of the velocity in 2-D and take tens of millions of nodes using higher-order methods and hundreds of millions for first-order methods \cite{Benson_AWR_2016}.  The properly upscaled RPT model used 20 initial $A$ (260 total) particles, and an ensemble of 100 simulations ran in minutes on a laptop PC.  {The volume-averaging upscaling method did not perform as well as the ensemble streamtube method for the $Pe$ and $Da$ values selected by {\em Kapoor et al.} \cite{Kapoor1997}.  The ensemble streamtube and RPT methods were essentially identical for this reactive scenario.  This suggests that the RPT method is performing an equivalent upscaling procedure automatically, because the particles experience the velocity perturbations as well as the local Green's function of mixing.  An interesting hypothesis is that the RPT method will succeed in upscaling regardless of the $Da, Pe$ regime because it does not discard any terms.}  If true, the particle method would be not only convenient, but theoretically preferred because, at any scale above the molecular scale in moving fluids, mixing, spreading, {and chemical kinetics} are completely different phenomena that should be simulated as such.  {Furthermore, the RPT method can accommodate any velocity field (with variability in time and space) and any reactions, whether fluid/fluid as done here, or fluid/solid \cite{Schmidt_fluid_solid}.}

\section{Acknowledgements}We thank the editor, Graham Sander, reviewers Olaf Cirpka, Giovanni Porta, and one anonymous reviewer for extremely helpful comments.  This material is based upon work supported by, or in part by, the US Army Research Office under Contract/Grant number W911NF-18-1-0338. The authors were also supported by the National Science Foundation under awards EAR-1417145, DMS-1211667, DMS-1614586, EAR-1351625, EAR-1417264, EAR-1446236, and CBET-1705770.

\section{Appendix: Upscaling Calculations}

Begin with the 2-D equations of transport and reaction at the micro-scale \eqref{eq:ADRE} with a total of three compounds $i=A,B,$ and $C$, with dispersion given by local diffusion (${\bf D} = D_{mol}$), reaction terms $R=-kc_Ac_B$ for $i=A=B$, and  $R=kc_Ac_B$ for $i=C$.  An equation for pseudo-species concentration $c_D=c_B-c_A$, given by subtracting the equation for $c_A$ from the same for $c_B$, is conservative due to the cancellation of their respective reaction terms. In particular, the resulting equation for $c_D$ is just \eqref{eq:ADRE} with $R=0$.
 
We may decompose concentrations and the velocity field into mean and fluctuation terms 
\beq\label{eq:fluctuation}
\begin{gathered}
c_i(t,x,y) = \overline{c}_i(t,x) + c'_i(t,x,y),\\
v_x(y) = \overline{v} + v_x'(y),
\end{gathered}
\eeq
where the overbar refers to the volume average across the $y$-direction and the prime to the zero-mean fluctuations about the average. 

\subsection{Two-equation volume-averaged closure}
We take a perturbative approach for upscaling by volume-averaging in the $y$-direction. In particular, we follow \cite{Porta2012} and \cite{Bolster2010} with an extension to include time-variable upscaled coefficients.  This approach discards a number of second- and higher-order moments in order to reach closure with two PDEs. 

For the conservative compound, place 
\eqref{eq:fluctuation} and $R=0$ into \eqref{eq:ADRE} 
\beq\label{eq:perturbADRE}
\frac{\partial \overline{c}_D }{\partial t}+ \frac{\partial c_D' }{\partial t} +(\overline {v} +v')( \frac{\partial \overline{c}_D }{\partial x}+\frac{\partial c_D' }{\partial x})=D_{mol}\frac{\partial^2 \overline{c}_D }{\partial x^2}  +D_{mol}\nabla^2 c_D'
\eeq
An average across $y$ gives
\beq\label{eq:averageADRE}
\frac{\partial \overline{c}_D }{\partial t}+ \overline {v} \frac{\partial \overline{c}_D }{\partial x}=D_{mol}\frac{\partial^2 \overline{c}_D }{\partial x^2} - \frac{\partial \overline{ v' c_D' }}{\partial x}
\eeq
%
The closure problem for $c_D'$ can be written by neglecting the second order term $\frac{\partial \overline{ v' c_D'}}{\partial x}$ and subsequently subtracting \eqref{eq:averageADRE} from \eqref{eq:perturbADRE}:
\beq\label{eq:perturb}
\frac{\partial c_D'}{\partial t}+ \overline{ v}\frac{\partial c_D'}{\partial x}+ v'  \frac{\partial \overline{c}_D }{\partial x}=D_{mol}\nabla^2 c_D'. 
\eeq
Following Porta \cite{Porta2012}, we assume that 
\beq
\label{eq:Porta_assumption}
c_D'=p\frac{\partial \overline{c}_D }{\partial x}
\eeq
which, when plugged into \eqref{eq:perturb} and assuming that transverse processes are most important, implies that $p= p(y,t)$ satisfies
\beq\label{eq:P2}
\frac{\partial p}{\partial t}+ v' =D_{mol}\frac{\partial^2 p}{\partial y^2} .
\eeq
The solution of this boundary-value problem is
\begin{eqnarray}
p(y,t)=-\int_0^t \int_0^b v'(\eta) G(y,t-\tau\vert \eta) d\eta d\tau
\end{eqnarray}
where $G$ is the transverse diffusion Green's function, namely
\begin{eqnarray}
G(y,t,\eta)=\frac{1}{b}+\frac{2}{b} \sum_{n=1}^\infty \cos\bigg(\frac{n\pi y}{b}\bigg) \cos\bigg(\frac{n\pi \eta}{b}\bigg) \exp \bigg(\frac{-D_{mol} n^2 \pi^2 t}{b^2}\bigg).
\end{eqnarray}
Thus,
\begin{eqnarray}
p(y,t)=\sum_{n=1}^\infty \frac{12 b^2 \overline{v}}{D_{mol}n^4 \pi^4} \cos\bigg(\frac{n\pi y}{b}\bigg) \bigg [1-\exp \bigg(\frac{-D_{mol} n^2 \pi^2 t}{b^2}\bigg)\bigg]
\end{eqnarray}
and this further implies
\beq
\frac{\partial \overline{ v' c_D' }}{\partial x}=- \overline{ v' p} \frac{\partial^2 \overline{C}_D }{\partial x^2}= - D_{mac}(t) \frac{\partial^2 \overline{ c}_D}{\partial x^2},
\eeq
where the upscaled hydrodynamic dispersion coefficient is given by
\begin{eqnarray}
D_{mac}(t) = 144 \frac{\overline{v} ^2 b^2}{\pi^6 D_{mol}} \sum_{n=1}^\infty ((1+(-1)^n) \frac{1}{n^6} \bigg [1-\exp \bigg(\frac{-D_{mol} n^2 \pi^2 t}{b^2}\bigg)\bigg].
\end{eqnarray}
We also note that for the values of the physical problem chosen here, the first non-zero term (i.e., $n=2$) provides a reasonable approximation
\begin{eqnarray}
D_{mac}(t) \approx  \frac{9\overline{v} ^2 b^2}{2\pi^6 D_{mol}} \bigg [1-\exp \bigg(\frac{-D_{mol} 4 \pi^2 t}{b^2}\bigg)\bigg]\approx4.75 \frac{mm^2}{s} (1-\exp(-t/25.3 s)).
\end{eqnarray}
\subsection{Reactive compounds}
Returning to the ADRE for the non-conservative compound $B$, averaging vertically yields the equation
\beq
\frac{\partial \overline { c}_B }{\partial t}+ \overline{v}\frac{\partial \overline{ c}_B }{\partial x}=D_{mol}\frac{\partial^2 \overline{ c}_B }{\partial x^2} - \frac{\partial \overline {v' c_B'}}{\partial x}-k\bigg[ \overline{ c}_B  ( \overline{ c}_B - \overline{ c}_D ) + \overline{ c'_B (c'_B-c'_D) }\bigg].
\eeq
To first order, the closure for $c_B'$ is 
\beq\label{eq:B'}
\frac{\partial c_B'}{\partial t}+ \overline{v}\frac{\partial c_B'}{\partial x}+ v' \frac{\partial  \overline{ c}_B}{\partial x}=D_{mol}\nabla^2 c_B' -k\bigg[ c'_B ( \overline{ c}_B -  \overline{ c}_D ) +  \overline{c}_B (c'_B-c'_D) \bigg].
\eeq
For fast reactions ($Da>>1$) the reaction term dominates, i.e. the right most term is larger than any other terms in the equation such that these can be neglected, meaning that the terms in the square parentheses sum to zero. As such, solving for $c_B'$ yields
\begin{eqnarray}\label{eq:fast}
c_B'= \frac{ \overline{ c}_B }{2 \overline{ c}_B - \overline{c}_D } c_D' = \frac{ \overline{c}_B }{2 \overline{c}_B - \overline{c}_D } p(y,t) \frac{\partial  \overline{ c}_D }{\partial x} = M(x,t) p(y,t) \frac{\partial  \overline{ c}_D }{\partial x}
\end{eqnarray}
where the mixing ratio is
$$M(x,t) = \frac{ \overline{ c}_B }{2 \overline{ c}_B - \overline{ c}_D } = \frac{ \overline{ c}_B }{ \overline{ c}_B +\overline{ c}_A } .$$
Thus, we find
\begin{eqnarray}
\frac{\partial \overline {v' c_B' }}{\partial x}=- \overline{ v' p } \frac{\partial }{\partial x} \left (M (x,t) \frac{\partial \overline{ c}_D }{\partial x} \right )= - D_{mac}(t) \frac{\partial }{\partial x} \left (M (x,t) \frac{\partial \overline{ c}_D }{\partial x} \right ),\end{eqnarray}
and using \eqref{eq:Porta_assumption},
\beq
\overline {c'_B (c'_B-c'_D) }= \overline{ M c_D^{'2} (M -1) }= M(M-1) \bigg( \frac{\partial \overline{c}_D}{\partial x}\bigg)^2 \overline {p^2 },
\eeq
where 
\begin{eqnarray}
\overline{ p^2 }=\sum_{n=1}^\infty \frac{144 \overline{v}^2 b^4}{D_{mol}^2 n^8 \pi^8} ((1+(-1)^n) \bigg[ 1-2\exp \bigg(\frac{-D_{mol} n^2 \pi^2 t}{b^2}\bigg) + \exp \bigg(\frac{-2 D_{mol} n^2 \pi^2 t}{b^2}\bigg) \bigg].
\end{eqnarray}

\noindent Therefore, our upscaled equations are:
\begin{eqnarray}\label{eq:Porta_start}
\frac{\partial \overline{ c}_D }{\partial t}+\overline{v}\frac{\partial \overline{ c}_D}{\partial x}=\bigg(D_{mol}+D_{mac}(t)\bigg) \frac{\partial^2 \overline{ c}_D }{\partial x^2},
\nonumber \\
\end{eqnarray}
\begin{eqnarray}\label{eq:upscaledB}
\frac{\partial \overline{ c}_B }{\partial t}+ \overline{v}\frac{\partial \overline{ c}_B }{\partial x}&=& D_{mol}\frac{\partial^2 \overline {c}_B}{\partial x^2} + D_{mac}(t) \frac{\partial }{\partial x} \left (M (x,t) \frac{\partial \overline {c}_D }{\partial x} \right )
\nonumber \\
& &- k \bigg( \overline {c}_B (\overline{c}_B - \overline{ c}_D ) +M(M-1) \bigg( \frac{\partial c_D}{\partial x}\bigg)^2  K(t) \bigg ),
\nonumber \\
\end{eqnarray}
where 
\begin{eqnarray}
 D_{mac}(t) = 144 \frac{\overline{v} ^2 b^2}{\pi^6 D_{mol}} \sum_{n=1}^\infty ((1+(-1)^n) \frac{1}{n^6} \bigg [1-\exp \bigg(\frac{-D_{mol} n^2 \pi^2 t}{b^2}\bigg)\bigg],
\end{eqnarray}
and 
\begin{eqnarray}\label{eq:Porta_end}
K(t)=\overline {p^2} =\sum_{n=1}^\infty \frac{144 \overline{v}^2 b^4}{D_{mol}^2 n^8 \pi^8} ((1+(-1)^n) \bigg[ 1-2\exp \bigg(\frac{-D_{mol} n^2 \pi^2 t}{b^2}\bigg) + \exp \bigg(\frac{-2 D_{mol} n^2 \pi^2 t}{b^2}\bigg) \bigg].
\end{eqnarray}

{Note that while we have generally followed \cite{Porta2012}, one difference does exist, which perhaps explains any possible discrepancies between our findings and theirs. In Porta's work they do not consider an explicit time dependent reaction coefficient as above. Later work, also by {\em Porta et al.} \cite{Porta_2016}, suggests that more complex closures can be used when coefficients are time dependent, resulting in a nonlocal integro-differential equation. However that system is significantly more computationally intensive than this closure. Additionally, the closure approximation was originally postulated strictly for cases where $Da\gg Pe$, although it was verified also for cases where $Da\gg1$ and $Pe$ and $Da$ had similar values as here.}

\subsection{{Streamtube mixing model}}

{Let us define global first, second, and second centered moments:}
{ \begin{eqnarray}
M_{1}(t) &=&\frac{1}{b}\int_0^b \int_{-\infty}^\infty x c(t,x,y) dx dy
\nonumber\\
M_{2}(t)&=&\frac{1}{b}\int_0^b \int_{-\infty}^\infty x^2 c(t,x,y) dx dy
\nonumber\\
K_2&=&M_2(t)-M^2_1(t).
\end{eqnarray}}
\noindent {The macroscopic dispersion coefficient is half the rate of change of the second centered global moment; i.e. }
{\begin{eqnarray}
D_{mac}=\frac{1}{2} \frac{d K_2}{dt}.
\end{eqnarray}}
\noindent {Similarly we can define local first, second, and second centered moments:}
{\begin{eqnarray}
m_{1}(y,t)&=&\int_{-\infty}^\infty x c(t,x,y) dx 
\nonumber\\
m_{2}(y,t)&=&\int_{-\infty}^\infty x^2 c(t,x,y) dx 
\nonumber\\
\kappa_2(y,t)&=&m_2(y,t)-m^2_1(y,t)
\end{eqnarray}}
{\noindent With these, we can define an alternative dispersion coefficient }
{\begin{eqnarray}
D_{\mathit{eff}}=\frac{1}{2} \frac{d }{dt} \bigg(\frac{1}{b} \int_0^b \kappa_2(y,t) dy\bigg),
\end{eqnarray}}
\noindent{which is a better measure of mixing than $D_{mac}$, which captures both mixing and spreading \cite{marco_mix_spread, Cirpka2000a}. Consider the difference between the first local and global moment, $p(y,t)=m_1(y,t)-M_1(t)$. It is straightforward to show that its governing equation and solution is the same as \eqref{eq:P2} such that}
{\begin{eqnarray}
p(y,t)=\sum_{n=1}^\infty \frac{12 b^2 \overline{v}}{D_{mol}n^4 \pi^4} \cos\bigg(\frac{n\pi y}{b}\bigg) \bigg [1-\exp \bigg(\frac{-D_{mol} n^2 \pi^2 t}{b^2}\bigg)\bigg]
\end{eqnarray}}
{\noindent The global and local second centered moments can be related by }
{\beq
K_2(t)=\frac{1}{b}\int_0^b \kappa_2(y,t) dy+\frac{1}{b} \int_0^b p^2(y,t) dy
\eeq}
\noindent {which means that the macroscopic and effective dispersion coefficients are related by}
{\beq
D_{mac}(t)-D_{\mathit{eff}}(t)=\frac{1}{2} \frac{d}{dt}\bigg(\frac{1}{b} \int_0^b p^2(y,t) dy\bigg) =\frac{1}{2} \frac{d \overline{p^2}}{dt}
\eeq}
\noindent  {Using \eqref{eq:P2} gives}
{\begin{eqnarray}\label{eq:Porta_end}
D_{mac}(t)&-&D_{\mathit{eff}}(t)
\nonumber
\\
&=& \sum_{n=1}^\infty \frac{144 \overline{v}^2 b^2}{D_{mol} n^6 \pi^6} ((1+(-1)^n) \bigg[ \exp \bigg(\frac{-D_{mol} n^2 \pi^2 t}{b^2}\bigg) - \exp \bigg(\frac{-2 D_{mol} n^2 \pi^2 t}{b^2}\bigg) \bigg]
\nonumber
\\
\end{eqnarray}}
\noindent{To leading order,}
{\begin{eqnarray}\label{eq:Porta_end}
D_{mac}(t)-D_{\mathit{eff}}(t)
\approx \frac{9 \overline{v}^2 b^2}{2 \pi^6 D_{mol} } \bigg[ \exp \bigg(\frac{-4 D_{mol} \pi^2 t}{b^2}\bigg) - \exp \bigg(\frac{-8 D_{mol} \pi^2 t}{b^2}\bigg) \bigg]
\end{eqnarray}}

\subsection{{Particle Numbers}}

{The number of particles used in the RPT model is based on the spatial covariance of concentrations.  Mixing in the Poiseuille system is dominated by transverse concentration gradients, so we may examine the transverse autocovariance and reactant segregation that develops almost immediately after the initial condition is distorted by the velocity field.  To our knowledge, expressions for the concentration statistics have not been developed for the reactive system.  Indeed, even the conservative system will experience dilution (and reduced total variance) as time gets large, so we look at the concentrations that develop early at a dimensionless time of $5t_A=t_D/20 =5$ s.  This ensures plenty of advective distortion but minimal dilution.}

\begin{figure}[h!]
 \centering
 \includegraphics[width=0.99\textwidth]{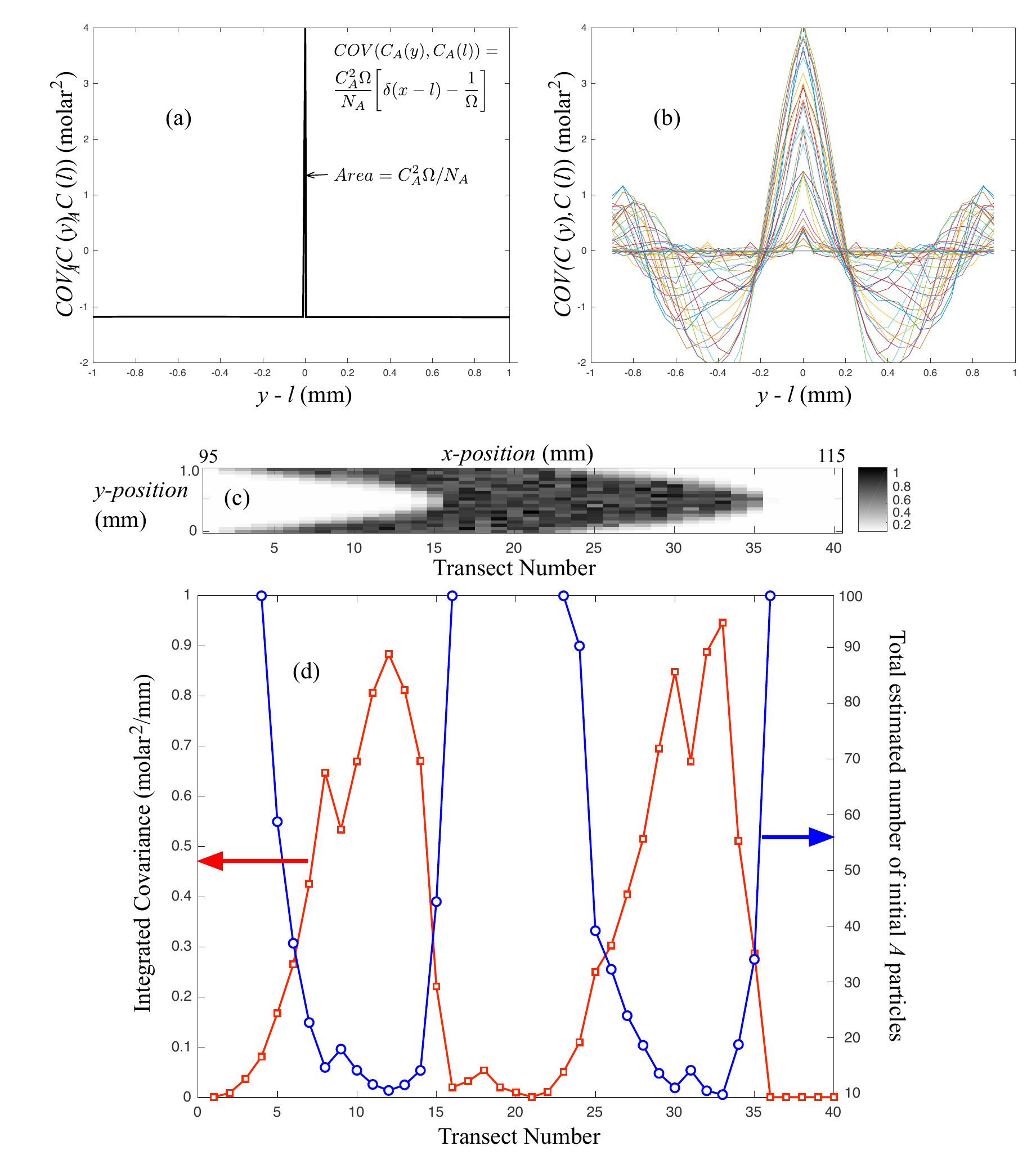}
 \caption{{(a) Plot of autocovariance function for uniformly randomly placed Dirac delta function particles in a domain of size 1 mm.  (b) Calculated $y$-direction autocovariance for nonreactive 260,000 particle 2-D model.  Functions calculated at 40 transects from $95\le x \le 115$ mm.  (c) Binned $A$ concentrations from nonreactive 260,000 particle 2-D model showing locations of $y$-transects used in subfigure (b).  (d) Red squares: Calculated integrated autocovariances (areas) from the 40 transects in the center ($y-l=0$), positive parts of the autocovariance functions in (b), and Blue circles: calculated total initial numbers of $A$ particles for upscaled 1-D models, based on calculated areas on the same plot.}}
 \label{fig:autocov}
 \end{figure}

{A 260,000 particle simulation was run with all parameters identical to the reactive simulation, except that the reaction rate coefficient was set to zero.  At a time of 5 s, the $A$ concentrations were binned (Fig. \ref{fig:autocov}c) and the covariance functions at each of 40 $y$-transects were calculated numerically (Fig. \ref{fig:autocov}b).  We are especially interested in the area corresponding to the positive covariance portion at small spatial separations in the center of these plots. A visual estimate of the ``extent'' of the $A$ concentrations is on the order of one-half pore width (Fig. \ref{fig:A_map}), which is the number we used in our uncalibrated simulations.  Of course the $A$ concentration covariance will be different along the length of the $A$ ``plume'', so we examine the variability here.}

{The autocovariance function (Fig. \ref{fig:autocov}a) for uniformly randomly placed Dirac-delta function particles placed in a finite domain was derived by {\em Schmidt et al.} \cite{Schmidt2017}:
\bl
\be \label{eq:COV}
COV(C_A(y),C_A(l))=\frac{N_Am_p^2}{\Omega}\biggl[ \delta(y-l)-\frac{1}{\Omega} \biggr]=C_Am_p\biggl[ \delta(y-l)-\frac{1}{\Omega} \biggr],
\ee
\el}
{where $C_A\approx C_A(t=0)$ is the initial, undiluted $A$ concentration, $N_A$ is the initial number of $A$ particles, $m_p$ is the mass of each particle, $\Omega$ is the extent of the domain (here 1 mm in the transverse direction), and $\delta$ is a Dirac-delta function.  Several studies have shown that equating the integral of this covariance to the integral of the real covariance function makes the particle model most closely match concentration evolution in real and numerical systems with concentration segregation (\cite{Paster_JCP,dong_awr,Ding_WRR, Schmidt2017,Bolster_mass}).  The fact that the delta particles have an atom of covariance at the origin (Fig. \ref{fig:autocov}a) is not a problem as they assume a Gaussian kernel shape when the mass transfer algorithm is applied.}

{Integrating the covariance function means that the near-origin integrated ``area'' for the delta particles is $C_Am_p$.  Because the particle mass is $m_p=C_A\Omega/N_A$, the area can be written $C_A^2\Omega/N_A$ (with units $[C^2 L]$). Equating this to the numerically estimated near-origin area (with units $[C^2 L]$) gives a particle density $N_A/\Omega=C_A^2/Area=1Mol^2/Area$.  This number is particles per mm in the $y$-direction.  The initial condition here is 10 mm in the $x$-direction, so the total number of initial $A$ particles is $10\times$ the density.  A plot of the near-origin areas and the resultant total number of $A$ particles (Fig. \ref{fig:autocov}d) shows that the number ranges from 11 to 40 in the high-mixing regions, with an average of 21, supporting our visual estimate of 20 for the initial condition. It may well be that a lower number would give an even better, i.e., slightly slower mixing, model, but we have not performed any calibration. }

\section*{References}

\bibliography{reaction_proposal}

\begin{thebibliography}{10}
\expandafter\ifx\csname url\endcsname\relax
  \def\url#1{\texttt{#1}}\fi
\expandafter\ifx\csname urlprefix\endcsname\relax\def\urlprefix{URL }\fi
\expandafter\ifx\csname href\endcsname\relax
  \def\href#1#2{#2} \def\path#1{#1}\fi

\bibitem{Benson_arbitrary}
D.~A. Benson, D.~Bolster,
  \href{http://dx.doi.org/10.1002/2016WR019368}{Arbitrarily complex chemical
  reactions on particles}, Water Resources Research 52~(11) (2016) 9190--9200.
\newblock \href {http://dx.doi.org/10.1002/2016WR019368}
  {\path{doi:10.1002/2016WR019368}}.
\newline\urlprefix\url{http://dx.doi.org/10.1002/2016WR019368}

\bibitem{Danckwerts1953}
P.~Danckwerts, Continuous flow systems: Distribution of residence times,
  Chemical Engineering Science 2~(1) (1953) 1--13.
\newblock \href {http://dx.doi.org/10.1016/0009-2509(53)80001-1}
  {\path{doi:10.1016/0009-2509(53)80001-1}}.

\bibitem{Nauman1983}
E.~Nauman, B.~Buffham, Mixing in continuous flow systems, Wiley, 1983.

\bibitem{Hill1976}
J.~C. Hill, Homogeneous turbulent mixing with chemical reaction, Annual Review
  of Fluid Mechanics 8 (1976) 135--161.

\bibitem{Molz_1988}
F.~Molz, M.~Widdowson, Internal inconsistencies in dispersion--dominated models
  that incorporate chemical and microbial kinetics, Water Resour. Res. 24
  (1988) 615--619.

\bibitem{Cirpka2000a}
O.~A. Cirpka, P.~K. Kitanidis,
  \href{https://agupubs.onlinelibrary.wiley.com/doi/abs/10.1029/1999WR900354}{Characterization
  of mixing and dilution in heterogeneous aquifers by means of local temporal
  moments}, Water Resources Research 36~(5) (2000) 1221--1236.
\newblock \href
  {http://arxiv.org/abs/https://agupubs.onlinelibrary.wiley.com/doi/pdf/10.1029/1999WR900354}
  {\path{arXiv:https://agupubs.onlinelibrary.wiley.com/doi/pdf/10.1029/1999WR900354}},
  \href {http://dx.doi.org/10.1029/1999WR900354}
  {\path{doi:10.1029/1999WR900354}}.
\newline\urlprefix\url{https://agupubs.onlinelibrary.wiley.com/doi/abs/10.1029/1999WR900354}

\bibitem{marco_mix_spread}
M.~Dentz, J.~Carrera, Mixing and spreading in stratified flow, Physics of
  Fluids 19 (2007) 017107.

\bibitem{Desimoni2005}
M.~De~Simoni, J.~Carrera, X.~S{\'a}nchez-Vila, A.~Guadagnini,
  \href{https://agupubs.onlinelibrary.wiley.com/doi/abs/10.1029/2005WR004056}{A
  procedure for the solution of multicomponent reactive transport problems},
  Water Resources Research 41~(11).
\newblock \href
  {http://arxiv.org/abs/https://agupubs.onlinelibrary.wiley.com/doi/pdf/10.1029/2005WR004056}
  {\path{arXiv:https://agupubs.onlinelibrary.wiley.com/doi/pdf/10.1029/2005WR004056}},
  \href {http://dx.doi.org/10.1029/2005WR004056}
  {\path{doi:10.1029/2005WR004056}}.
\newline\urlprefix\url{https://agupubs.onlinelibrary.wiley.com/doi/abs/10.1029/2005WR004056}

\bibitem{donado}
L.~Donado, X.~S\'anchez-Vila, M.~Dentz, J.~Carrera, D.~Bolster, Multicomponent
  reactive transport in multicontinuum media, Water Resour. Res. 45 (2009)
  W11402, doi:10.1029/2008WR006823.

\bibitem{Rezaei}
M.~Rezaei, E.~Sanz, E.~Raeisi, C.~Ayora, E.~V{\'a}zquez-Su{\~n}{\'e},
  J.~Carrera,
  \href{http://www.sciencedirect.com/science/article/pii/S0022169405000739}{Reactive
  transport modeling of calcite dissolution in the fresh-salt water mixing
  zone}, Journal of Hydrology 311~(1) (2005) 282 -- 298.
\newblock \href
  {http://dx.doi.org/https://doi.org/10.1016/j.jhydrol.2004.12.017}
  {\path{doi:https://doi.org/10.1016/j.jhydrol.2004.12.017}}.
\newline\urlprefix\url{http://www.sciencedirect.com/science/article/pii/S0022169405000739}

\bibitem{Dentz2011}
M.~Dentz, T.~L. Borgne, A.~Englert, B.~Bijeljic,
  \href{http://www.sciencedirect.com/science/article/pii/S0169772210000495}{Mixing,
  spreading and reaction in heterogeneous media: A brief review}, Journal of
  Contaminant Hydrology 120-121 (2011) 1 -- 17, reactive Transport in the
  Subsurface: Mixing, Spreading and Reaction in Heterogeneous Media.
\newblock \href
  {http://dx.doi.org/https://doi.org/10.1016/j.jconhyd.2010.05.002}
  {\path{doi:https://doi.org/10.1016/j.jconhyd.2010.05.002}}.
\newline\urlprefix\url{http://www.sciencedirect.com/science/article/pii/S0169772210000495}

\bibitem{Lehwald2012}
A.~Lehwald, G.~Janiga, D.~Th{\'e}venin, K.~Z{\"a}hringer, Simultaneous
  investigation of macro- and micro-mixing in a static mixer, Chemical
  Engineering Science 79~(0) (2012) 8--18.
\newblock \href {http://dx.doi.org/10.1016/j.ces.2012.05.026}
  {\path{doi:10.1016/j.ces.2012.05.026}}.

\bibitem{tanguy_prl}
T.~Le~Borgne, M.~Dentz, E.~Villermaux, Stretching, coalescence, and mixing in
  porous media, Physical Review Letters 110~(20) (2013) 204501.
\newblock \href {http://dx.doi.org/10.1103/PhysRevLett.110.204501}
  {\path{doi:10.1103/PhysRevLett.110.204501}}.

\bibitem{Tanguy_GRL2014}
T.~{Le Borgne}, T.~R. Ginn, M.~Dentz,
  \href{http://dx.doi.org/10.1002/2014GL062038}{Impact of fluid deformation on
  mixing-induced chemical reactions in heterogeneous flows}, Geophysical
  Research Letters 41~(22) (2014) 7898--7906.
\newblock \href {http://dx.doi.org/10.1002/2014GL062038}
  {\path{doi:10.1002/2014GL062038}}.
\newline\urlprefix\url{http://dx.doi.org/10.1002/2014GL062038}

\bibitem{Ding_WRR}
D.~Ding, D.~A. Benson, D.~Fern{\'a}ndez-Garcia, C.~V. Henri, D.~W. Hyndman,
  M.~S. Phanikumar, D.~Bolster,
  \href{http://dx.doi.org/10.1002/2017WR021103}{Elimination of the reaction
  rate ``scale effect'': Application of the lagrangian reactive
  particle-tracking method to simulate mixing-limited, field-scale
  biodegradation at the {S}choolcraft ({MI}, {USA}) site}, Water Resources
  Research~(doi:10.1002/2017WR021103).
\newblock \href {http://dx.doi.org/10.1002/2017WR021103}
  {\path{doi:10.1002/2017WR021103}}.
\newline\urlprefix\url{http://dx.doi.org/10.1002/2017WR021103}

\bibitem{Schmidt_accuracy}
M.~J. Schmidt, S.~D. Pankavich, D.~A. Benson,
  \href{https://www.sciencedirect.com/science/article/pii/S0309170818301830}{On
  the accuracy of simulating mixing by random-walk particle-based mass-transfer
  algorithms}, Advances in Water Resources (2018) --\href
  {http://dx.doi.org/https://doi.org/10.1016/j.advwatres.2018.05.003}
  {\path{doi:https://doi.org/10.1016/j.advwatres.2018.05.003}}.
\newline\urlprefix\url{https://www.sciencedirect.com/science/article/pii/S0309170818301830}

\bibitem{Schmidt_fluid_solid}
M.~J. Schmidt, S.~D. Pankavich, A.~Navarre-Sitchler, D.~A. Benson, A lagrangian
  method for reactive transport with solid/aqueous chemical phase interaction,
  https://arxiv.org/abs/1805.06072.

\bibitem{White_Peterson_1990}
A.~F. White, M.~L. Peterson,
  \href{https://pubs.acs.org/doi/abs/10.1021/bk-1990-0416.ch035}{Role of
  Reactive-Surface-Area Characterization in Geochemical Kinetic Models}, ACS
  Publications, 1990, Ch.~35, pp. 461--475.
\newblock \href
  {http://arxiv.org/abs/https://pubs.acs.org/doi/pdf/10.1021/bk-1990-0416.ch035}
  {\path{arXiv:https://pubs.acs.org/doi/pdf/10.1021/bk-1990-0416.ch035}}, \href
  {http://dx.doi.org/10.1021/bk-1990-0416.ch035}
  {\path{doi:10.1021/bk-1990-0416.ch035}}.
\newline\urlprefix\url{https://pubs.acs.org/doi/abs/10.1021/bk-1990-0416.ch035}

\bibitem{Brantley_book}
S.~L. Brantley, J.~D. Kubicki, A.~F. White (Eds.), Kinetics of Water-Rock
  Interaction, Springer, 2008.

\bibitem{Kapoor1997}
V.~Kapoor, L.~W. Gelhar, F.~Miralles-Wilhelm,
  \href{http://dx.doi.org/10.1029/96WR03687}{Bimolecular second-order reactions
  in spatially varying flows: Segregation induced scale-dependent
  transformation rates}, Water Resour. Res. 33~(4) (1997) 527--536.
\newline\urlprefix\url{http://dx.doi.org/10.1029/96WR03687}

\bibitem{Kapoor1998a}
V.~Kapoor, C.~T. Jafvert, D.~A. Lyn, Experimental study of a bimolecular
  reaction in {P}oiseuille flow, Water Resour. Res. 34~(8) (1998) 1997--2004.
\newblock \href {http://dx.doi.org/10.1029/98WR01649}
  {\path{doi:10.1029/98WR01649}}.

\bibitem{Taylor1953}
G.~Taylor,
  \href{http://rspa.royalsocietypublishing.org/content/219/1137/186}{Dispersion
  of soluble matter in solvent flowing slowly through a tube}, Proceedings of
  the Royal Society of London A: Mathematical, Physical and Engineering
  Sciences 219~(1137) (1953) 186--203.
\newblock \href
  {http://arxiv.org/abs/http://rspa.royalsocietypublishing.org/content/219/1137/186.full.pdf}
  {\path{arXiv:http://rspa.royalsocietypublishing.org/content/219/1137/186.full.pdf}},
  \href {http://dx.doi.org/10.1098/rspa.1953.0139}
  {\path{doi:10.1098/rspa.1953.0139}}.
\newline\urlprefix\url{http://rspa.royalsocietypublishing.org/content/219/1137/186}

\bibitem{Bolster2010}
D.~Bolster, F.~J. Vald{\'e}s-Parada, T.~LeBorgne, M.~Dentz, J.~Carrera, Mixing
  in confined stratified aquifers, Journal of contaminant hydrology 120 (2011)
  198--212.

\bibitem{Dentz2000}
M.~Dentz, H.~Kinzelbach, S.~Attinger, W.~Kinzelbach,
  \href{https://agupubs.onlinelibrary.wiley.com/doi/abs/10.1029/2000WR900162}{Temporal
  behavior of a solute cloud in a heterogeneous porous medium: 1. point-like
  injection}, Water Resources Research 36~(12) (2000) 3591--3604.
\newblock \href
  {http://arxiv.org/abs/https://agupubs.onlinelibrary.wiley.com/doi/pdf/10.1029/2000WR900162}
  {\path{arXiv:https://agupubs.onlinelibrary.wiley.com/doi/pdf/10.1029/2000WR900162}},
  \href {http://dx.doi.org/10.1029/2000WR900162}
  {\path{doi:10.1029/2000WR900162}}.
\newline\urlprefix\url{https://agupubs.onlinelibrary.wiley.com/doi/abs/10.1029/2000WR900162}

\bibitem{Cirpka2000b}
O.~A. Cirpka, P.~K. Kitanidis,
  \href{https://agupubs.onlinelibrary.wiley.com/doi/abs/10.1029/1999WR900355}{An
  advective-dispersive stream tube approach for the transfer of
  conservative-tracer data to reactive transport}, Water Resources Research
  36~(5) (2000) 1209--1220.
\newblock \href
  {http://arxiv.org/abs/https://agupubs.onlinelibrary.wiley.com/doi/pdf/10.1029/1999WR900355}
  {\path{arXiv:https://agupubs.onlinelibrary.wiley.com/doi/pdf/10.1029/1999WR900355}},
  \href {http://dx.doi.org/10.1029/1999WR900355}
  {\path{doi:10.1029/1999WR900355}}.
\newline\urlprefix\url{https://agupubs.onlinelibrary.wiley.com/doi/abs/10.1029/1999WR900355}

\bibitem{Bolster2012AWR}
D.~Bolster, P.~de~Anna, D.~A. Benson, A.~M. Tartakovsky, Incomplete mixing and
  reactions with fractional dispersion, Advances in Water Resources 37 (2012)
  86--93.

\bibitem{Tartakovsky2012}
A.~M. Tartakovsky, P.~de~Anna, T.~Le~Borgne, A.~Balter, D.~Bolster, Effect of
  spatial concentration fluctuations on non-linear reactions in
  diffusion-reaction systems, Water Resour. Res. 48 (2012) W02526.

\bibitem{Paster_JCP}
A.~Paster, D.~Bolster, D.~A. Benson,
  \href{http://www.sciencedirect.com/science/article/pii/S0021999114000473}{Connecting
  the dots: {S}emi-analytical and random walk numerical solutions of the
  diffusion--reaction equation with stochastic initial conditions}, Journal of
  Computational Physics 263 (2014) 91 -- 112.
\newblock \href {http://dx.doi.org/https://doi.org/10.1016/j.jcp.2014.01.020}
  {\path{doi:https://doi.org/10.1016/j.jcp.2014.01.020}}.
\newline\urlprefix\url{http://www.sciencedirect.com/science/article/pii/S0021999114000473}

\bibitem{Schmidt2017}
M.~J. Schmidt, S.~Pankavich, D.~A. Benson, A kernel-based {L}agrangian method
  for imperfectly-mixed chemical reactions, Journal of Computational Physics
  336 (2017) 288 -- 307.
\newblock \href {http://dx.doi.org/http://doi.org/10.1016/j.jcp.2017.02.012}
  {\path{doi:http://doi.org/10.1016/j.jcp.2017.02.012}}.

\bibitem{Porta2012}
G.~M. Porta, J.-F. Thovert, M.~Riva, A.~Guadagnini, P.~M. Adler, Microscale
  simulation and numerical upscaling of a reactive flow in a plane channel,
  Phys. Rev. E 86~(3) (2012) 036102--.
\newblock \href {http://dx.doi.org/10.1103/PhysRevE.86.036102}
  {\path{doi:10.1103/PhysRevE.86.036102}}.

\bibitem{Porta2013}
G.~M. Porta, S.~Chaynikov, J.-F. Thovert, M.~Riva, A.~Guadagnini, P.~M. Adler,
  Numerical investigation of pore and continuum scale formulations of
  bimolecular reactive transport in porous media, Advances in Water Resources
  62, Part B (2013) 243--253.
\newblock \href {http://dx.doi.org/10.1016/j.advwatres.2013.09.007}
  {\path{doi:10.1016/j.advwatres.2013.09.007}}.

\bibitem{Raje}
D.~Raje, V.~Kapoor, {Experimental study of bimolecular reaction kinetics in
  porous media}, Environ. Sci. \& Tech. {34}~({7}) ({2000}) {1234--1239}.

\bibitem{gramling}
C.~Gramling, C.~Harvey, L.~Meigs, Reactive transport in porous media: A
  comparison of model prediction with laboratory visualization, Environmental
  Science \& Technology 36~({11}) (2002) 2508--2514.
\newblock \href {http://dx.doi.org/10.1021/es0157144}
  {\path{doi:10.1021/es0157144}}.

\bibitem{Edery2009}
Y.~Edery, H.~Scher, B.~Berkowitz,
  \href{http://dx.doi.org/10.1029/2008GL036381}{Modeling bimolecular reactions
  and transport in porous media}, Geophys. Res. Lett. 36~(2) (2009) L02407.
\newline\urlprefix\url{http://dx.doi.org/10.1029/2008GL036381}

\bibitem{Edery2010}
Y.~Edery, H.~Scher, B.~Berkowitz,
  \href{http://dx.doi.org/10.1029/2009WR009017}{Particle tracking model of
  bimolecular reactive transport in porous media}, Water Resour. Res. 46~(7)
  (2010) W07524.
\newline\urlprefix\url{http://dx.doi.org/10.1029/2009WR009017}

\bibitem{sanchez-vila}
X.~S\'anchez-Vila, D.~Fern\`andez-Garcia, A.~Guadagnini, Interpretation of
  column experiments of transport of solutes undergoing an irreversible
  bimolecular reaction using a continuum approximation, Water Resour. Res. 46
  (2010) W12510.

\bibitem{Zhang_PRE_react}
Y.~Zhang, C.~Papelis, Particle-tracking simulation of fractional
  diffusion-reaction processes, Phys. Rev. E 84 (2011) 066704.

\bibitem{dong_awr}
D.~Ding, D.~Benson, A.~Paster, D.~Bolster, Modeling bimolecular reactions and
  transport in porous media via particle tracking, Advances in Water Resources
  53 (2012) 56--65.
\newblock \href {http://dx.doi.org/10.1016/j.advwatres.2012.11.001}
  {\path{doi:10.1016/j.advwatres.2012.11.001}}.

\bibitem{Bolster_mass}
D.~Bolster, A.~Paster, D.~A. Benson,
  \href{http://dx.doi.org/10.1002/2015WR018310}{A particle number conserving
  {L}agrangian method for mixing-driven reactive transport}, Water Resources
  Research 52~(2) (2016) 1518--1527.
\newblock \href {http://dx.doi.org/10.1002/2015WR018310}
  {\path{doi:10.1002/2015WR018310}}.
\newline\urlprefix\url{http://dx.doi.org/10.1002/2015WR018310}

\bibitem{Benson_react}
D.~A. Benson, M.~M. Meerschaert,
  \href{http://dx.doi.org/10.1029/2008WR007111}{Simulation of chemical reaction
  via particle tracking: Diffusion-limited versus thermodynamic rate-limited
  regimes}, Water Resour. Res. 44 (2008) W12201.
\newblock \href {http://dx.doi.org/10.1029/2008WR007111}
  {\path{doi:10.1029/2008WR007111}}.
\newline\urlprefix\url{http://dx.doi.org/10.1029/2008WR007111}

\bibitem{Benson_AWR_2016}
D.~A. Benson, T.~Aquino, D.~Bolster, N.~Engdahl, C.~V. Henri,
  D.~Fern{\`a}ndez-Garcia, A comparison of {E}ulerian and {L}agrangian
  transport and non-linear reaction algorithms, Advances in Water Resources 99
  (2017) 15 -- 37.
\newblock \href
  {http://dx.doi.org/http://doi.org/10.1016/j.advwatres.2016.11.003}
  {\path{doi:http://doi.org/10.1016/j.advwatres.2016.11.003}}.

\bibitem{labolle}
E.~M. Labolle, G.~E. Fogg, A.~F.~B. Tompson, Random-walk simulation of
  transport in heterogeneous porous media: {L}ocal mass-conservation problem
  and implementation methods, Water Resour. Res. 32~(3) (1996) 583--593.

\bibitem{Salamon_2006}
P.~Salamon, D.~Fern{\`a}ndez-Garcia, J.~J. G{\'o}mez-Hern{\'a}ndez,
  \href{http://www.sciencedirect.com/science/article/pii/S0169772206000957}{A
  review and numerical assessment of the random walk particle tracking method},
  Journal of Contaminant Hydrology 87~(3--4) (2006) 277 -- 305.
\newblock \href
  {http://dx.doi.org/http://dx.doi.org/10.1016/j.jconhyd.2006.05.005}
  {\path{doi:http://dx.doi.org/10.1016/j.jconhyd.2006.05.005}}.
\newline\urlprefix\url{http://www.sciencedirect.com/science/article/pii/S0169772206000957}

\bibitem{Tartakovsky_Hagan}
A.~M. Tartakovsky, D.~{Barajas-Solano}, Persistent incomplete mixing in
  reactive flows, https://arxiv.org/abs/1803.06693v1.

\bibitem{Porta_2016}
G.~Porta, G.~Ceriotti, J.-F. Thovert, Comparative assessment of continuum-scale
  models of bimolecular reactive transport in porous media under pre-asymptotic
  conditions, Journal of Contaminant Hydrology 185--186 (2016) 1 -- 13.
\newblock \href {http://dx.doi.org/10.1016/j.jconhyd.2015.12.003}
  {\path{doi:10.1016/j.jconhyd.2015.12.003}}.

\end{thebibliography}

\end{document}